\documentclass[pdftex,english,10pt,twocolumn]{article}
\usepackage{palatino}
\usepackage{mathpazo}
\usepackage[english]{babel}
\usepackage[utf8]{inputenc}
\usepackage[margin=0.8in]{geometry}
\usepackage{framed}
\usepackage[small]{titlesec}
\usepackage{hyperref}
\usepackage{xcolor}
\usepackage{mathtools}

\usepackage{graphicx}
\usepackage{subfig}
\usepackage[labelfont=bf,labelsep=period,justification=raggedright]{caption}

\definecolor{darkblue}{rgb}{0.0,0.0,0.75}
\hypersetup{colorlinks,breaklinks,
            linkcolor=darkblue,urlcolor=darkblue,
            anchorcolor=darkblue,citecolor=darkblue}

\usepackage[autostyle]{csquotes}
\usepackage[sort&compress,square,numbers]{natbib}

\RequirePackage{authblk}

\usepackage{seqsplit}
\newcommand{\wangbar}{\rule[-3pt]{1.5pt}{11pt}}
\usepackage{stfloats}
\usepackage[activate={true,nocompatibility},final,tracking=true,kerning=true,spacing=true,factor=1100,stretch=10,shrink=10]{microtype}
\microtypecontext{spacing=nonfrench}
\usepackage[normalem]{ulem}
\usepackage{fancyhdr}

\usepackage{amsthm,amsmath,amssymb,graphicx,pdflscape,multirow}
\usepackage{array}
\usepackage{xcolor,colortbl}
\usepackage{multicol}
\usepackage{multirow}
\usepackage{wrapfig,lipsum,booktabs}
\usepackage{fancybox}
\usepackage{setspace}
\usepackage{adjustbox}
\usepackage{enumitem}
\usepackage{bbm}
\usepackage{arydshln}
\usepackage{hyperref}
\usepackage{soul}

\newcolumntype{L}[1]{>{\raggedright\let\newline\\\arraybackslash\hspace{0pt}}m{#1}}
\newcolumntype{C}[1]{>{\centering\let\newline\\\arraybackslash\hspace{0pt}}m{#1}}
\newcolumntype{R}[1]{>{\raggedleft\let\newline\\\arraybackslash\hspace{0pt}}m{#1}}

\definecolor{Gray}{gray}{0.9}
\definecolor{DarkGray}{gray}{0.75}

\usepackage{url} 
\usepackage{csquotes}

\begin{document}

\title{\bf Open Case Studies: Statistics and Data Science Education through Real-World Applications}

\author[1]{\bf Carrie Wright} 
\author[1]{\bf Qier Meng} 
\author[2]{\bf Michael R. Breshock} 
\author[2]{\bf Lyla Atta} 
\author[1]{\bf Margaret A.\ Taub } 
\author[1]{\bf Leah R. Jager } 
\author[1,3]{\bf  John Muschelli } 
\author[1,*]{\bf Stephanie C.\ Hicks} 

\affil[1]{Department of Biostatistics, Johns Hopkins Bloomberg School of Public Health}
\affil[2]{Department of Biomedical Engineering, Johns Hopkins University}
\affil[3]{Streamline Data Science}
\affil[*]{Correspondence to \nolinkurl{shicks19@jhu.edu}}

\makeatletter
\let\Title\@title
\let\Author\@author
\makeatother

\date{}
\maketitle

\thispagestyle{fancy}

\begin{abstract}

\noindent With unprecedented and growing interest in data science education, there are limited educator materials that provide meaningful opportunities for learners to practice {\it statistical thinking}, as defined by \citet{wild1999}, with messy data addressing real-world challenges. 
As a solution, \citet{nolan1999} advocated for bringing applications to the forefront in undergraduate statistics curriculum with the use of in-depth {\it case studies} to encourage and develop statistical thinking in the classroom. 
Limitations to this approach include the significant time investment required to develop a case study -- namely, to select a motivating question and to create an illustrative data analysis -- and the domain expertise needed.
As a result, case studies based on realistic challenges, not toy examples, are scarce. 
To address this, we developed the Open Case Studies (\href{https://www.opencasestudies.org}{opencasestudies.org}) project, which offers a new statistical and data science education case study model. This educational resource provides self-contained, multimodal, peer-reviewed, and open-source guides (or case studies) from real-world examples for active experiences of complete data analyses. 
We developed an educator's guide describing how to most effectively use the case studies, how to modify and adapt components of the case studies in the classroom, and how to contribute new case studies. (\href{https://www.opencasestudies.org/OCS_Guide}{opencasestudies.org/OCS\_Guide}). \\


\end{abstract}

\section{Introduction}
\label{sec:intro}

A major challenge in the practice of teaching data science and statistics is the limited  availability of courses and course materials that provide meaningful opportunities for students to practice and apply {\it statistical thinking}, as defined by \citet{wild1999}, with messy data addressing real-world challenges across diverse context domains. To address this problem, \citet{nolan1999} presented a model for developing {\it case studies} (also known as `labs') for use in undergraduate statistics courses with a specific goal to ``encourage and develop statistical thinking". Specifically, the model calls for each case study to be:

\begin{displayquote}
``a substantial exercise with nontrivial solutions that leave room for different analyses, and for it to be a central part of the course. The lab should offer motivation and a framework for studying theoretical statistics, and it should give students experience with how statistics can be used to answer scientific questions. An important goal of this approach is to encourage and develop statistical thinking while imparting knowledge in mathematical statistics.'' \cite{nolan1999}
\end{displayquote}

In 2018, \citet{hicks2018} stated that one of their five principles for teaching data science was to ``organize the course around a set of diverse case studies'' based on the model by \citet{nolan1999}, with a goal of practicing statistical thinking and bringing real-world applications into the classroom. Case studies are also being used in the classroom across a diverse set of fields, including statistics \cite{weinberg_casestudies_2000, schafer_casestudies_2003, khachatryan_v_2017, rivera_incorporating_2019, donoghue_casestudies_2021}, evolutionary biology \cite{reyes_casestudies_2022}, engineering \cite{romero_teaching_1995}, and environmental science \cite{theobold_casestudies_2021}. 

However, there are several limiting factors to scaling up the use of case studies.  First, the process of selecting motivating questions \cite{arnold_what_2021}, finding real-world and motivating data \cite{neumann_using_2013, Donoho2017}, describing the context around the data \cite{Wood_2018, committee_on_envisioning_the_data_science_discipline_the_undergraduate_perspective_data_2018}, and preparing diverse didactic data analyses requires a large initial investment in time and effort \cite{hicks2018}. Second, the individuals who are most primed to develop effective and insightful case studies are practitioner-instructors \cite{KrossGuo2019}, or practicing applied statisticians and data scientists, who teach and practice in a field-specific context. For these individuals, successfully constructing a diverse set of case studies across a wide range of contextual topics may require collaboration with individuals in other disciplines; this can be hard without protected time and effort from their academic institutions \cite{Waller2018}. Third, while there are rich repositories of data sets \cite{rivera_incorporating_2019}, there are few collections of associated data analyses that show how the data can be used to demonstrate fundamental data science and statistical concepts, potentially with unexpected outcomes \cite{peng_diagnosing_2021}. This is especially true for complex and messy data, where analysis decisions must go beyond what can be summarized in a brief summary about the data, such as a README file \cite{vilhuber_teaching_2022, dogucu_tools_2022}. These challenges have resulted in a scarcity of case studies based on real-world challenges instead of simple toy examples. Moreover, many data repositories have different recommended processing and analysis of subsets of data, which are commonly used as "the" analysis, without proper discussion of alternative choices along the research pathway. 

To address these challenges, we developed an open-source educational resource, the Open Case Studies (OCS) project (\href{https://www.opencasestudies.org}{opencasestudies.org}). This resource contains in-depth, self-contained, multimodal, and peer-reviewed experiential guides (or case studies) that demonstrate illustrative data analyses covering a diverse range of statistical and data science topics to teach learners how to effectively derive knowledge from data. 
These guides can be used by instructors to bring applications to the forefront in the classroom or they can be used by independent learners outside of the classroom. Finally, we developed an educator's guide describing how to most effectively use the case studies, how to modify and adapt components of the case studies in the classroom, and how to contribute new case studies. (\href{https://www.opencasestudies.org/OCS_Guide}{opencasestudies.org/OCS\_Guide}). 

The rest of the manuscript is as follows. First, we provide an overview and discuss individual components of the Open Case Studies model (\textbf{Section~\ref{sec:puttinginpractice}}), a new model that extends the \cite{nolan1999} case studies model. Second, we describe the Open Case Studies educational resource (\textbf{Section~\ref{sec:ocs-resource}}). Third, we give guidance based on our experience about how others can create their own case studies (\textbf{Section~\ref{sec:ocs-guide}}), including how to create interactive case studies. We conclude with a summary about the utility of such case studies inside and outside of the classroom (\textbf{Section~\ref{sec:discussion}}).


\section{Putting OCS model into practice}
\label{sec:puttinginpractice}

\subsection{An overview of the Open Case Studies model}

The case-studies model described by \citet{nolan1999} divides each case study into five main components: (i) introduction, (ii) data description, (iii) background, (iv) investigations, and (v) theory, with an optional section for advanced analyses or related theoretical material. In our Open Case Studies (OCS) model, we expand upon these components to thirteen components. Table~\ref{tab-map-structure} describes the components of the OCS model as well as the mapping between our model and the original model of \citet{nolan1999}. 

\begin{table*}[ht!]
\footnotesize
\centering
\begin{framed}
\begin{tabular}{ |p{3.3cm}|p{5cm}||p{2.3cm}|p{4.3cm}|}
\hline
 \multicolumn{4}{|c|}{\textbf{Mapping of components between two case study models}} \\
\hline
\multicolumn{2}{|c||}{ \cellcolor{DarkGray} Open Case Studies model} & \multicolumn{2}{c|}{\cellcolor{DarkGray} Case-study model of \citet{nolan1999}} \\
 \hline
 \hline
 \textbf{Component} & \textbf{Description} & \textbf{Component} & \textbf{Description} \\
 \hline
 \hline
\cellcolor{Gray} \multirow{2}{*}{1. Motivation} & \cellcolor{Gray} Motivating figure and text at the start of the case study &  &  \\
 2. Main questions & Scientific question(s) &  \multirow{-3}{*}{Introduction} & \multirow{-3}{4.2cm}{Describes context of scientific question and motivation}  \\
\hline
\hline
\cellcolor{Gray} \multirow{2}{*}{3. Learning objectives}   &  \cellcolor{Gray}  Both data science and statistics learning objectives &  &   \\
4. Context & Context of question(s) or data &  &  \\
\cellcolor{Gray} \multirow{2}{*}{5. Limitations} &   \cellcolor{Gray} Any limitations in case study or with data used & \multirow{-3}{*}{Background} & \multirow{-3}{4.2cm}{Information to put question in context using non-technical language}  \\
\hline
\hline
 \multirow{2}{*}{6. What are the data?} & Summary of where the data came from and what the data contain & \multirow{2}{*}{Data description} & Documentation for data collected to address the question  \\
\hline
\hline
\cellcolor{Gray} 7. Data import & \cellcolor{Gray} Analyses for importing data  & &  \\ 
8. Data wrangling and exploration  & Analyses for wrangling and exploring the data  & & \\ 
\cellcolor{Gray} 9. Data visualization &  \cellcolor{Gray} Analyses for data visualization & \multirow{-4}{*}{Investigations}   & \multirow{-4}{4.2cm}{Suggestions for answering the question (varies in difficulty)} \\ 
\hline
\hline
\multirow{3}{*}{10. Data analysis}  & Analyses containing statistical concepts and methods to answer question(s) & \multirow{3}{*}{Theory} & Describes relevant statistical concepts and methodologies to answer the question  \\
\hline
\hline
\cellcolor{Gray} 11. Summary &  \cellcolor{Gray} Summary of results  & &  \\ 
12. Suggested homework  & Question(s) to explore further & & \\ 
\cellcolor{Gray} 13. Additional information   &  \cellcolor{Gray} Helpful links or packages used & \multirow{-3}{3cm}{Extended material (optional)} & \multirow{-3}{4.2cm}{Describes advanced analyses or related theoretical material} \\      
 \hline
\end{tabular}
\caption{\textbf{Components of an Open Case Study \wangbar} Descriptions of the components of our Open Case Studies model (left) and their mapping to the components of the case studies model proposed by \citet{nolan1999} (right). We note that the model from \citet{nolan1999} orders `Data description' before `Background'. However, Background is listed first here to more easily map to our Open Case Studies model.}
\label{tab-map-structure}
\end{framed}
\end{table*}

We highlight that while the structures of the two case-study models are similar, our OCS model has a different purpose than the one proposed by \citet{nolan1999}. Briefly, \citet{nolan1999} designed case studies to be either (i) used in open-ended discussions in lecture or (ii) used as open-ended lab exercises where students do extensive analyses outside of class and write reports containing their observations and solutions. In both applications, the case studies are designed to be open-ended; the background may be initially discussed in class or as part of an assignment, but students work independently or in a group to create their own solutions and summarize their own findings in a full-length report to answer the original question. In contrast, we made a design choice to build case studies that are full-length, in-depth experiential guides that walk learners through the entire process of data analysis, with an emphasis on computing \cite{nolan_computing}, starting from a motivating question and ending with a summary of the results. Our goal is for educators either to directly use an entire case study in the classroom or to adapt a subset of the material for their use. For example, an educator can choose to show the solutions provided in the case study, show a different solution, or leave the discussion open-ended. Our reasoning for providing full-length guides is that it is typically easier for an educator to remove or modify material instead of creating it from scratch. In this way, we aim to reach a broader audience than just educators in a classroom, as any learner interested in a particular topic can walk through the case study to see an example of a complete data analysis. In addition, this method is particularly helpful for instructors who may not feel confident creating an analysis from scratch, especially if it is outside their main area of expertise, as our case studies built with domain experts and are peer-reviewed.

\subsection{Components of the Open Case Studies model}

\begin{figure*}[!ht]
\centering
\begin{framed}
\includegraphics[width=0.90\textwidth]{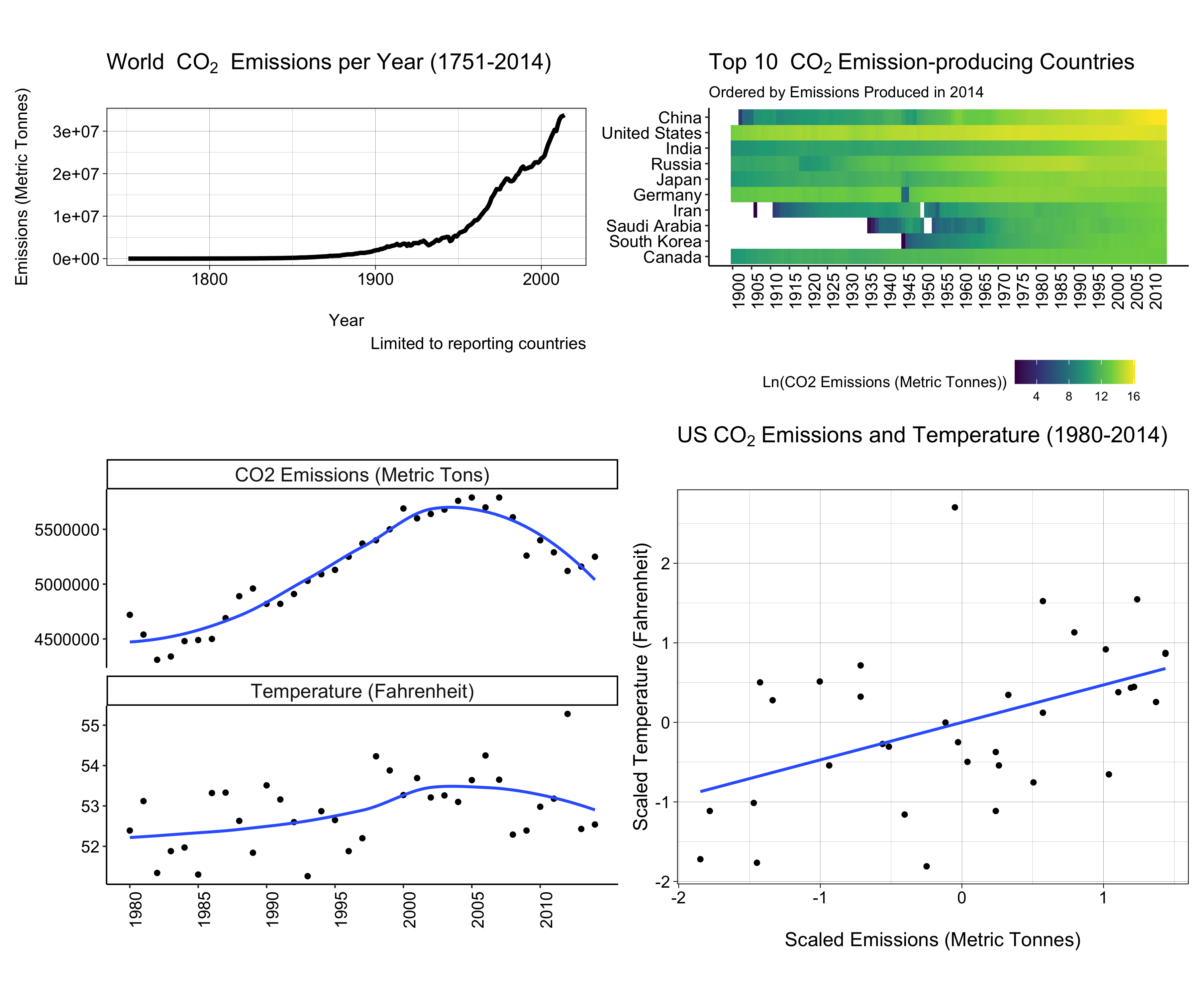}
 \caption{
 {\small \textbf{Example of a motivating figure in the ``Exploring CO$_2$ emissions across time'' case study \wangbar} 
 The complete case study can be found at (\href{https://www.opencasestudies.org/ocs-bp-co2-emissions}{opencasestudies.org/ocs-bp-co2-emissions}). 
 \textbf{Top row}: Line plot showing the increase in CO$_2$ emissions over time (left). Longitudinal heatmap plot highlighting that the US has been one of the top emission producing countries historically and currently (right). \textbf{Bottom row}: Scatter plots showing the trends between CO$_2$ emissions and temperature across time. 
 } 
 } 
  \label{fig-motivating-figure}
\end{framed}
\end{figure*}

We will describe the thirteen individual components of our Open Case Studies model (\textbf{Table~\ref{tab-map-structure}}) using one case study as an example. Currently all of our case studies showcase how to use the R statistical programming language \cite{R_2021} for data analyses, although other programming languages could be used with our model. Here, we use the ``Exploring CO$_2$ emissions across time'' case study (\href{https://www.opencasestudies.org/ocs-bp-co2-emissions}{opencasestudies.org/ocs-bp-co2-emissions}), which explores global and country level carbon dioxide (CO$_2$) emissions from the 1700s to 2014 (\textbf{Figure~\ref{fig-motivating-figure}}). This case study also investigates how CO$_2$ emission rates may relate to increasing temperatures and increasing rates of natural disasters in the United States (US). We also describe four other case studies (\textbf{Table~\ref{tab-exOCS}}) and give example topics covered in all case studies (\textbf{Table~\ref{tab-topics}}). \\ 

\begin{table*}[ht!]
\begin{framed}
\scriptsize
\centering
\begin{tabular}{|C{1.6cm}|C{3cm}|C{2.8cm}|C{1.5cm}|C{2.7cm}|C{2.7cm}|}
\hline
 \multicolumn{6}{|c|}{\textbf{Example case studies in the OCS resource}}\\
\hline
\cellcolor{DarkGray} Topic & \cellcolor{DarkGray} Question(s) & \cellcolor{DarkGray} Data source(s)& \cellcolor{DarkGray}Raw data & \cellcolor{DarkGray}Data science skills	& \cellcolor{DarkGray}Statistical concepts\\
\hline
\end{tabular}
\begin{tabular}{|L{1.6cm}|L{3cm}|L{2.8cm}|L{1.5cm}|L{2.7cm}|L{2.7cm}|}
Air Pollution [\href{https://www.opencasestudies.org/ocs-bp-air-pollution}{html}] & Can we predict annual fine particulate air pollution concentrations using predictors such as population density, urbanization, and satellite data? & 
Gravimetric EPA air pollution data (from 2008) and predictor data from NASA, the US Census, and NCHS & Single curated CSV file & 
tidymodels, correlation visualizations, geospatial visualizations & machine learning, linear regression, random forest\\
\hline
\cellcolor{Gray} Vaping [\href{https://www.opencasestudies.org/ocs-bp-vaping-case-study}{html}] &
\cellcolor{Gray}How has tobacco / nicotine product use by American youths changed since 2015? Is there a relationship between e-cigarette / vaping use and other tobacco / nicotine product use? & 
\cellcolor{Gray}NYTS 2015-2019 survey data & 
\cellcolor{Gray}Excel files and codebooks for each year & 
\cellcolor{Gray}importing Excel files, importing multiple files efficiently, merging data, writing functions, functional programming, longitudinal visualizations & 
\cellcolor{Gray}survey weighting, logistic regression with survey weighting, longitudinal data, codebooks \\
\hline
CO$_2$ Emissions [\href{https://www.opencasestudies.org/ocs-bp-co2-emissions}{html}] & How have global CO$_2$ emission rates changed over time? In particular for the US, and how does the US compare to other countries? // 
Are CO$_2$ emissions in the US, global temperatures, and natural disaster rates in the US associated? & 
CO$_2$ emissions (from 1751-2019, GDP and energy use data from gapminder. US temperature and disaster data form the NOAA & 
XLSX and CSV files & importing data from Excel files and CSV files, data joining, longitudinal data visualizations, plots with text and labels & correlation coefficient, relationship between correlation and linear regression, correlation vs. causation \\
\hline
\cellcolor{Gray}US School Shootings [\href{https://www.opencasestudies.org/ocs-bp-school-shootings-dashboard}{html}] & 
\cellcolor{Gray}What has been the yearly rate of school shootings and where in the country have they occurred in the last 50 years (from January 1970 to June 2020)? & 
\cellcolor{Gray}Open-source K-12 school shooting database (1970-2020) &
\cellcolor{Gray}single CSV file, Google sheets & 
\cellcolor{Gray}importing Google sheets, date formats, geocoding, interactive tables, R Markdown,  maps, interactive dashboards & 
\cellcolor{Gray}calculating percentages for data with missing values\\
\hline
\end{tabular}
\caption{\textbf{Description of four example case studies in the OCS resource \wangbar} This table shows the topics covered in four individual case studies, as well as information about the raw data. EPA = the US Environmental Protection Agency, NASA = National Aeronautics and Space Administration, and NCHS = the National Center for Health Statistics, NYTS = the National Youth Tobacco Survey, NOAA = National Oceanic and Atmospheric Administration. CO$_2$ emission data obtained from gapminder was originally from the World Bank. School Shooting data was obtained from the Center for Homeland Defense and Security at the at the Naval Postgraduate School (NPS).}
\label{tab-exOCS}
\end{framed}
\end{table*}

\noindent \textbf{1. Motivation}. Each case study begins with a motivating data visualization. This idea originated from Dr. Mine Çetinkaya-Rundel's talk entitled `Let Them Eat Cake (First)!', presented at the Harvard University Statistics Department's 2018 David K. Pickard Memorial Lecture  \cite{mine2018}. She argues that, similar to a recipe book about baking cakes, showing a learner a visualization first can be motivating and give learners a sense of what they will be doing. This practice of showing a visualization at the start of a data analysis and then showing learners the code for how to produce the data visualization enables the learners to have a better sense of the final product and can be motivating to learn the more challenging concepts needed to make the visualization. 

The motivating figure from the CO$_2$ emissions case study (\textbf{Figure \ref{fig-motivating-figure}}) is reproduced here. In the case studies, we also include text explaining the motivation for the case study. Our case studies are often motivated by a recent report or publication investigating a specific scientific question. In this section, we explain why the topic is of interest and define any terms that are needed to understand the main questions of interest (described in the next section). \\

\noindent \textbf{2. Main questions}. In this section, we highlight and explicitly state a precise set of scientific question(s) or problem(s) before beginning the analysis \cite{ratan_formulation_2019}. When the case study is motivated by a previous publication, these questions may not be exactly the same as what was originally investigated in the paper or report. For example, a case study may only investigate a small subset of the results presented in the report or publication. Alternatively, a case study may not investigate the same question(s) at all, but rather use the data from the report or publication to demonstrate a specific data science or statistics learning objective. 
This framework also reiterates that many problems have a set of questions prior to analysis; finding an answer and engineering the question post-doc is not recommended. Data exploration is a large component of the analysis framework and is shown in case studies, but OCS impresses thoughtful questions should be determined prior to analysis.

In the CO$_2$ emissions case study, the scientific questions are: 

\begin{enumerate}[noitemsep]
    \item How have global CO$_2$ emission rates changed over time? In particular for the US, and how does the US compare to other countries?
    \item Are CO$_2$ emissions in the US, global temperatures, and natural disaster rates in the US associated?
\end{enumerate}


\noindent \textbf{3. Learning objectives}. Each case study consists of a set of didactic learning objectives. We categorize each objective as related to either (i) data science or (ii) statistics where the latter are concepts traditionally taught in a statistics curriculum such as linear regression, multiple testing correction, significance and the former are concepts often appearing outside of a traditional statistics course, such as re-coding data values, scraping data from a website, or creating a dashboard for a data set. Other categories could be considered depending on the purpose of the case study. This separation also allows for educators to adapt the material to other computational frameworks and languages other than R, such as Python.

We include these learning objectives for three reasons: (i) to help educators select a case study that meets the objectives they want to teach and (ii) to help learners select a case study that demonstrates what they want to learn, and (iii) to provide both educators and learners with a clear understanding about the goals of a particular case study. For example, a study of the use of learning objectives in an undergraduate science course found that students find learning objectives helpful for narrowing and organizing their studying \cite{osueke_how_2018}.

For the CO$_2$ emissions case study, we designed the case study around the following learning objectives: \\

\noindent \underline{(i) Data Science Learning Objectives:}

\begin{itemize}[noitemsep]
    \item Importing data from various types of Excel files and CSV files
    \item Apply action verbs in \texttt{dplyr} \cite{pkg-dplyr} for data wrangling
    \item How to pivot between ``long'' and ``wide'' data sets
    \item Joining together multiple datasets using \texttt{dplyr}
    \item How to create effective longitudinal data visualizations with \texttt{ggplot2} \cite{ggplot2}
    \item How to add text, color, and labels to \texttt{ggplot2}  plots
    \item How to create faceted \texttt{ggplot2} plots
\end{itemize}

\noindent \underline{(ii) Statistical Learning Objectives:}

\begin{itemize}[noitemsep]
    \item Correlation coefficient as a summary statistic
    \item Relationship between correlation, linear regression
    \item Correlation is not causation
\end{itemize}

In addition, by stating these objectives within the case studies, students may begin identify how they can apply these concepts for future analyses. 
Finally, we provide an interactive search table of learning objectives on the Open Case Studies website (\href{https://www.opencasestudies.org}{opencasestudies.org}) to make it easier to find a case study that would demonstrate a particular technique, method, or concept that an instructor or learner might be interested in. \\

\noindent \textbf{4. Context}. The context section provides background information needed to understand the context of the question(s) of interest and the data that will be used to answer the questions \cite{Wood_2018, committee_on_envisioning_the_data_science_discipline_the_undergraduate_perspective_data_2018}.  This may include information from the publication on which the case study is based, but also additional background literature. For an example from public health, the case study may describe what is currently known (or not known) about the health impact of the topic. This serves to demonstrate how the specific question(s) fit into a larger scientific context. 

For the CO$_2$ case study, the context section includes a discussion of the potential impacts of climate change on human health, an overview of the likely progression of warming in the coming years, and potential impacts on other components of the environment such as ocean acidity and rainfall quantities. \\

\noindent \textbf{5. Limitations}. In addition to the motivation and context for each case study, it is important to formally describe limitations of the analysis presented as it provides important context for the educator or learner \cite{rivera_incorporating_2019}. Examples of limitations include (i) limitations due to the available data, such as the use of surrogate variables or indicators, (ii) limitations in the methods used, such as annual average estimates for quantities that are likely to vary daily or monthly, and (iii) selection biases due to sampling of observed data. A key concept in data science is that the conclusions from an analysis can only be as good as the data that go into it and the methods used to analyze them, so presenting these limitations provides a valuable learning opportunity.

In the CO$_2$ case study, we describe limitations about how the data are incomplete because only certain countries reported CO$_2$ emissions for certain years. We describe how additional emissions were also produced by countries that are not included in the data. This helps the learners to understand that while the data will help us understand CO$_2$ emissions, it will not provide the full picture. \\ 

\noindent \textbf{6. What are the data?}  To provide transparency about the data sources, we describe where and how the raw data were obtained and used in the case study. If the data are obtained from a website, survey or report, and where possible, we also describe how the data were originally collected. We typically describe what the variables are in each dataset later in the case study to better match the experience of the learners discovering the data after they import and explore it. 

The data sources for the CO$_2$ case study are from Gapminder (\href{https://www.gapminder.org}{gapminder.org}) (originally from the World Bank) and the United States National Oceanic and Atmospheric Administration. In the case study, we present a table with the different data sources and a brief description of each one, including sources to cite. \\

\noindent \textbf{7. Data import}. Next, we describe the steps and give the code required to read the raw data into the analysis environment. Currently, all of our case studies describe analyses in the R programming language. In some cases, importing the raw data is fairly straightforward, and this section is quite short.  Other case studies have longer and more involved data import sections that involve scraping data from a PDF, accessing data using an Application Programming Interface (API), or writing functions to efficiently access data from multiple files with the same format. Importantly, we describe all of our use of code in the case studies in a literate programming way \cite{knuth_literate_1984}, meaning that we describe each step in a way that can be understandable by learners.

Since the data for the CO$_2$ case study are stored in Excel and comma-separated-variable (CSV) files, we use standard data import functions \texttt{read\_excel()} from the \texttt{readxl} package and \texttt{read\_csv()} from the \texttt{readr} R package \cite{pkg-readr} to import our data.\\

\noindent \textbf{8. Data wrangling and exploration}. Typically one of the longest sections for many of our case studies is the wrangling section, which describes all of the steps required to take the imported raw data and get it into a state that is ready for analysis and creating visualizations. We also demonstrate how to perform exploratory data analysis \cite{Tukey1962}. 

For example in the CO$_2$ case study, the raw data needs to be converted from a ``wide'' to ``long'' format  so that each country-year observation is in a single row.
After wrangling the data from each source, we demonstrate how to join together data sets from different sources by matching on country-year combinations. Ultimately, we create one large data set containing all the variables we want to use for our analysis (in the columns) with one record for each country-year combination (in the rows).\\

\noindent \textbf{9. Data visualization}. We show both simple and complex data visualizations to explore and demonstrate a variety of graphical design choices, including plot type and other aesthetic choices to best show the types of variables of interest. In addition, most case studies describe how to facet or combine plots together so that all the major findings of the case study are illustrated in a single data visualization. 

In the CO$_2$ case study, we create data visualizations for a subset of the variables. For example, we use line plots to visualize how CO$_2$ emissions, in metric tons, have varied over time globally (\textbf{Figure~\ref{fig-motivating-figure}}). We go into detail around coloring and labeling the lines, zooming in and out on the time-scale axis, as well as including informative plot titles and axis labels. We demonstrate that when looking at CO$_2$ emissions from different countries across time, special consideration for labeling is required. We show that a heat map or tile plot does a great job of illustrating top country differences in a less overwhelming manner (\textbf{Figure~\ref{fig-motivating-figure}}). We also demonstrate the utility of faceted plots to simultaneously visualize more variables over time. We also show how to start looking for associations or trends in the data through scatter plots with smoothed lines or linear regression lines added.\\

\noindent \textbf{10. Data analysis}. Our case studies are intended to introduce how a particular statistical test or data science technique might be implemented and interpreted to answer the scientific question(s) of interest. However, we walk the learner through an unexpected outcome and how we diagnosed it \cite{peng_diagnosing_2021}. We provide background information about statistical concepts and how these concepts apply to our example analysis.

The main topic of the analysis section for our CO$_2$ emissions Open Case Study is correlation and how correlation is related to linear regression. We discuss background information such as a description about what summary statistics are, what the correlation coefficient is, and how the correlation coefficient is mathematically calculated. We also describe the limitations of correlation analysis and how correlation does not imply causation. We demonstrate how to implement assessments of correlation and how to interpret the results.\\

\noindent \textbf{11. Summary}. In this section we provide a summary figure that visually indicates some of the major findings of the case study. The goal of this visualization is to demonstrate how to communicate the results of the analysis to a broader audience \cite{khachatryan_v_2017}. This often involves combining plots and adding annotations. This summary figure is the motivating figure used at the beginning of the case study. Along with this figure, we provide a synopsis of the case study in which the motivation, context, and scientific questions are restated and summarized, while the major steps of wrangling, data exploration, and analysis are described. The main findings of the analysis are discussed, with emphasis on what these findings might indicate for the larger context of the scientific question, in addition to what still remains unknown.

In the CO$_2$ emissions Open Case Study, the summary figure (\textbf{Figure~\ref{fig-motivating-figure}}) combines several of the plots from the case study together to demonstrate the major findings. The synopsis recaps what data we worked with (CO$_2$ emissions for some countries from 1751- 2014) and what we have shown in the analysis, including touching on the learning objectives outlined at the beginning.    We give a simpler explanation about the statistical concepts that were discussed in the analysis section, in this case about correlation and regression. We discuss more about what we were able to answer or not answer in terms of the questions of interest. We describe how we discovered a dramatic increase in global CO$_2$ emissions over time and that some countries appear to be especially responsible. We discuss that although the data suggests a relationship between temperature and CO$_2$ emissions in the US, there are many other important factors to consider based on what we know about climate change. These include: the influence of CO$_2$ emissions from other countries in the atmosphere, the influence of other greenhouse gases, the fact that the already existing CO$_2$ in the atmosphere continues to trap heat for many years, and the fact that heat trapped in the ocean due to previous emissions causes delayed changes in surface temperatures.  We also point out what the results of our analysis might mean for how we should consider mitigating climate change effects and how warming temperatures may impact society in the future. \\ 

\noindent \textbf{12. Suggested homework}. Each case study suggests a homework activity for students to try on their own. These activities typically require the students to use the skills that they have learned on a new data set or to expand the analysis to evaluate another subset of the data. Students may also be asked to make visualizations based on these analyses.

The suggested homework for the CO$_2$ emissions Open Case Study are to:
\begin{itemize}[noitemsep]
    \item Create a plot with labels showing the countries with the lowest CO$_2$ emission levels.
    \item Plot CO$_2$ emissions and other variables (e.g. energy use) on a scatter plot, calculate the Pearson’s correlation coefficient, and discuss results.
\end{itemize}

These suggestions would require learners to practice their visualization an analytic skills to further investigate the data with less guidance. \\

\noindent \textbf{13. Additional information}. This section includes additional information about the broader scientific topic of the case study, the methods used to analyze the data, and the specific data sets used in the analysis.  Information is provided as links to external online resources such as blog posts, scientific articles, scientific reports, and educational websites.  We also provide links to documentation about the R packages used, as well as the specific package versions that were used. We also link to information about the specific subject-matter experts who contributed to the development of the case study.

The CO$_2$ emissions Open Case Study includes links to resources for learning more about the various R packages used in the case study (such as \texttt{here} \cite{here},
\texttt{readxl} \cite{readxl}, \texttt{readr} \cite{pkg-readr}, \texttt{dplyr} \cite{pkg-dplyr}, \texttt{magrittr} \cite{magrittr}, \texttt{stringr} \cite{stringr}, \texttt{purrr} \cite{purrr}, \texttt{tidyr} \cite{tidyr}, \texttt{tibble} \cite{tibble}, \texttt{forcats} \cite{forcats}, ggplot2 \cite{ggplot2}, \texttt{directlabels} \cite{directlabels}, \texttt{ggrepel} \cite{ggrepel}, \texttt{broom} \cite{broom}, \texttt{patchwork} \cite{patchwork}) and how they were used, as well as information about the statistical topics touched on, including correlation, regression and time series analysis. These go beyond some of the material presented in the case study, to help point instructors or learners to additional resources for topics of interest.\\

\section{The OCS educational resource}
\label{sec:ocs-resource}

The OCS resource can be found online (\href{https://www.opencasestudies.org}{opencasestudies.org}). In addition, we created an educator's guide describing how to most effectively use the case studies, how to modify and adapt components of the case studies in the classroom, and how to contribute new case studies. (\href{https://www.opencasestudies.org/OCS_Guide}{opencasestudies.org/OCS\_Guide}). 

\subsection{\textbf{Open Case Study website and search tool}}
\label{sec:ocs-resource-website}

\begin{figure*}[hb!]
\centering
\begin{framed}
\includegraphics[width=0.95\textwidth]{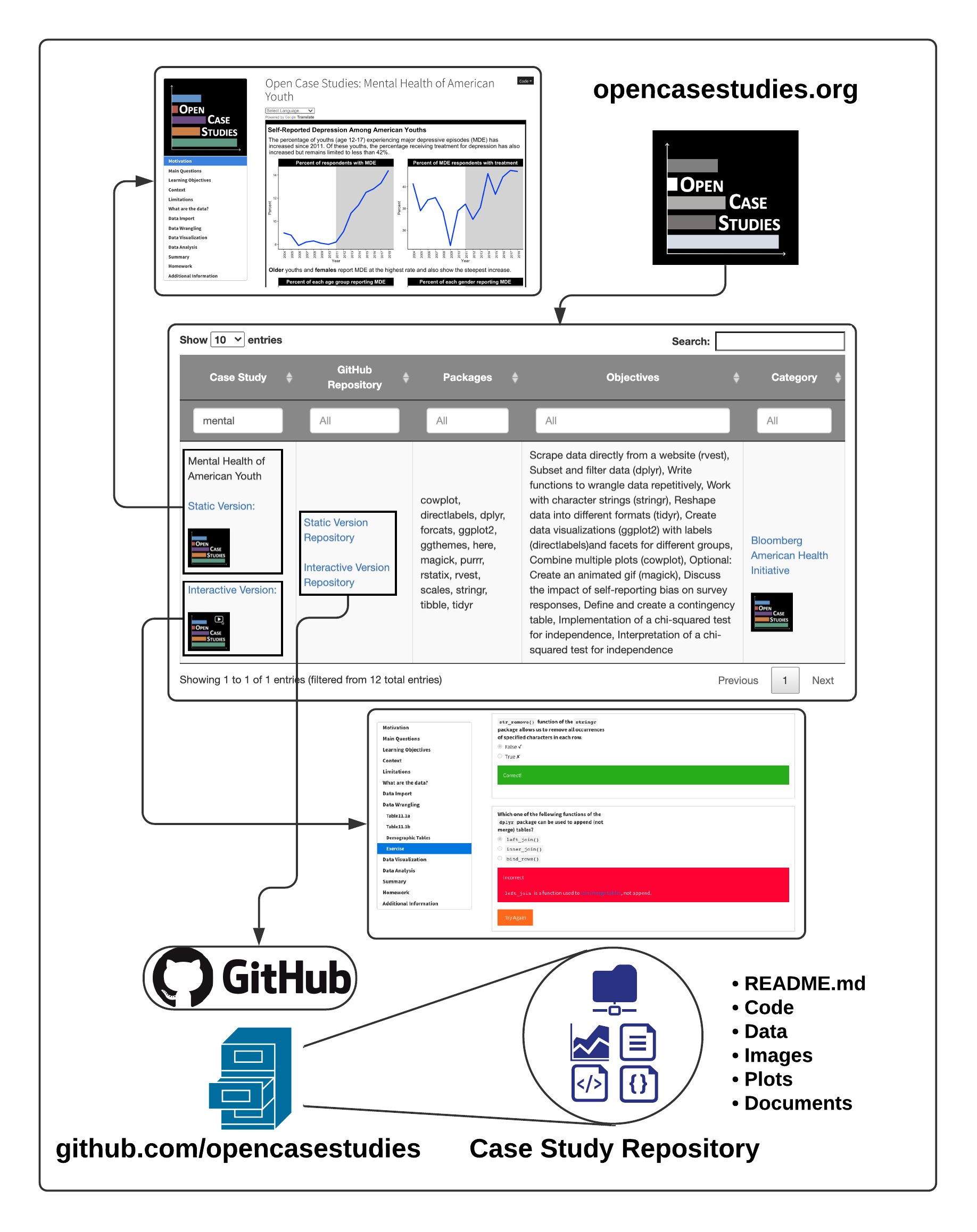}
 \caption{
 {\small \textbf{An overview of the OCS educational resource \wangbar} The Open Case Studies website contains a searchable database of all available case studies. Users can search by case study name, R packages used, learning objectives, and category. Each case study links to a website with a rendered version of the entire analysis and to the Github repository. The Github repository hosts the online lesson and all of the related code, data, image, plot, and document files needed to follow along or conduct new analyses. Some case studies now have interactive versions that include live quizzes and coding tutorials.} 
 } 
  \label{OCS_Access}
\end{framed}
\end{figure*}

Our case study resource is hosted on our Open Case Studies (OCS) website (\textbf{Figure~\ref{OCS_Access}}). To navigate the case studies, we provide an interactive search table, built using the \texttt{DT} package \cite{DT}, that allows those interested to search through our case studies by topic, statistical learning objective, data science learning objective, and R packages demonstrated. This table includes links to the code and data for each case study, as well as a links to websites that are rendered versions of each case study where the entire analysis can be read in full. 

\subsection{\textbf{Open Case Studies on GitHub}} 
\label{sec:ocs-resource-github}

The code and data for each case study are hosted in a GitHub repository (\textbf{Figure~\ref{OCS_Access}}). 
Our case studies are built in R Markdown, allowing text, images, and gifs that describe the context and data analytic process to be interspersed with code chunks that show the actual code used in the analysis \cite{knuth_literate_1984}. We developed these prior to the release of the quarto publishing system (\href{https://quarto.org/quarto}{quarto.org/quarto}). These case studies are then ``knit'' into rendered html-formatted files using GitHub actions \cite{noauthorfeaturesnodate} for continuous integration and deployment. By continuous integration, we mean that changes are tracked and a history of the code from various authors is saved to a single main version \cite{shahin_continuous_2017} using Git and GitHub. By continuous deployment, we mean that the website versions of the case studies are automatically rendered and available to the public once a new version is established on GitHub. These website versions of the case studies are also hosted on GitHub. Currently our case studies are all written using the R programming language, however our current format could be extended to support tutorials using other programming languages as well.
Our case studies have a table of contents that allows instructors and learners to easily navigate from section to section, so that they can focus on the materials most useful for their needs. In addition, each case study starts with a graphic or plot that describes the basic findings of the case study. Each case study is organized with the same basic structure so that learners can navigate case studies more easily, and see patterns across case studies on how analysis is performed (\textbf{Figure~\ref{ocs_str}}).

\subsection{\textbf{Open Case Study file structure}}
\label{sec:ocs-resource-structure}

\begin{figure*}[ht!]
\centering
\begin{framed}
\includegraphics[width=0.75\textwidth]{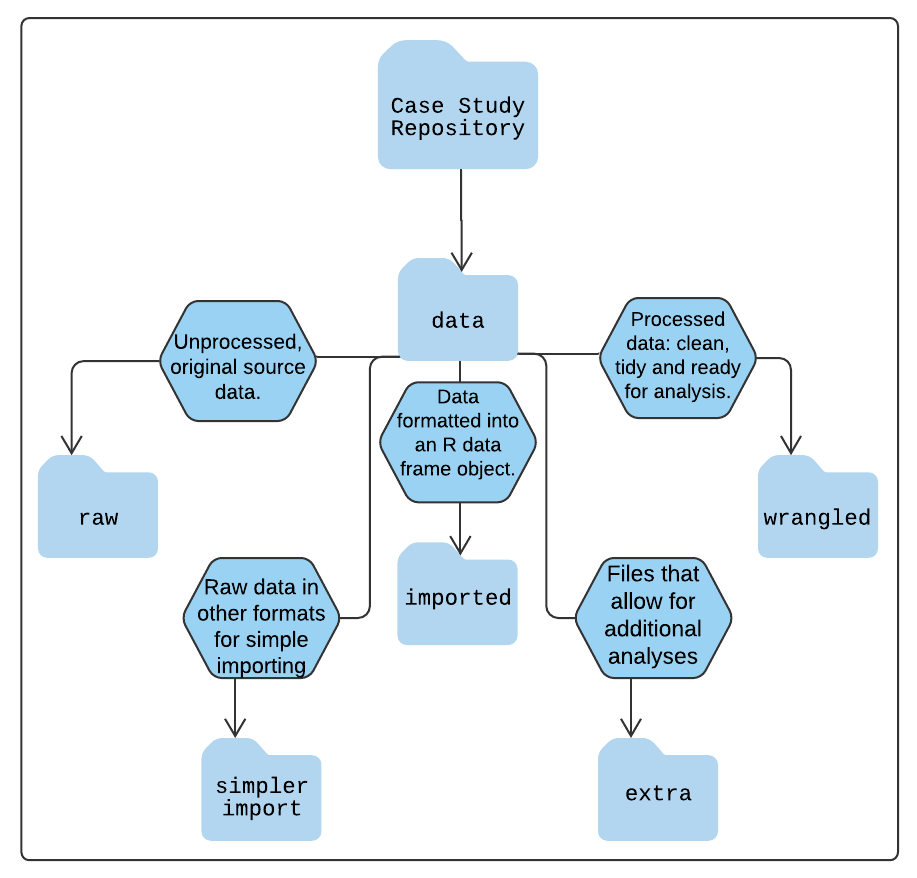}
 \caption{
 {\small \textbf{An overview of the \texttt{data} file structure on GitHub \wangbar} A tree illustrating the repository data directory structure. Each bubble describes the type of data files that can be found in the sub-folders.} 
 } 
  \label{Data_str}
\end{framed}
\end{figure*}

Each case-study repository has a similar file structure, with a data directory containing both raw data and versions of the data in various processed forms to allow instructors/learners to modularize the case studies for their own  purposes (\textbf{Figure \ref{Data_str}}). For example, an instructor could skip the data import and wrangling sections of the case study and focus on the visualizations and analysis pieces using a fully cleaned data set.  To support this modular style of instruction, each case study includes commands at the beginning of each section that imports the data in the final state of the previous section.  These different stages of the data are organized in a data folder with five categories: raw, imported, wrangled, simpler import, and extra.  The raw data directory includes files in their original unaltered condition and in the original file format from the original data source (in some cases raw files are CSV files, Excel files, PDFs among other file formats). The imported data directory includes files containing the data in a format that is directly compatible with R, such as RData files which are often abbreviated as Rda. The wrangled data directory also includes an RData file that contains a clean and tidy version of the data that has been pre-processed and is ready for analysis, as well as csv files for instructors that wish to demonstrate a simpler version of data import. The simpler import folder contains raw files that have been converted to CSV file format or other formats that can be more easily imported into R. The extra data folder contains data files that allow for individuals to conduct analyses beyond what was done in the case study (the file format for these extra files varies). Each repository also contains a README file \cite{vilhuber_teaching_2022, dogucu_tools_2022} that explains the modular aspect of the case study, as well as other information about how to use the case study for educational purposes (\textbf{Table~\ref{tab-mod}}).

\subsection{\textbf{Interactive elements in Open Case Studies.}} 
\label{sec:ocs-resource-interactive}

\label{sec:int-ocs} To make our case studies more experiential, we have introduced interactive elements including quizzes and coding exercises using the \texttt{learnr} \cite{learnr} and \texttt{gradethis} \cite{gradethis} packages. 

We include a mix of multiple choice questions and coding exercises in each case study.  Coding exercises are embedded throughout the content of the case studies and give students a chance to write code for a specific step in the analysis. The answers to these exercises (the code/output used in the case study to complete these steps) is then hidden in a click-to-expand section right after the exercise window. Students can compare their own code and output with these answers. We also create exercise subsections at the end of the main sections of the case study. These exercise subsections include both multiple choice questions and coding exercises. Students can use them to test their understanding of the content in each section. All multiple-choice questions provide real-time feedback, giving hints after wrong answers and allowing students to retry the questions if they submitted a wrong answer. For most of the coding exercises, hints and/or solutions are available. With the help of the \texttt{gradethis} package, some of these coding problems also provide real-time feedback after students submit their code.

\section{Building your own case studies}
\label{sec:ocs-guide}

For educators interested in constructing their own case studies, in this next section, we describe our recommendations for the process based on our experiences and challenges throughout this project. We also describe these recommendations in our Educator's guide (\href{https://www.opencasestudies.org/OCS_Guide}{opencasestudies.org/OCS\_Guide}). 

\subsection{Identifying questions and data for case studies}

The process of choosing data sources and questions of interest is arguably the most important part of constructing a case study.  We can either identify an interesting and publicly available data set and then ask a timely and engaging question about a topic related to the data, or we can identify an interesting question and then work to find publicly available data to answer this question. This process of linking a question to publicly available data often involves a bit of trial and error and reshaping of the question while keeping in mind and potentially adjusting what the case study is meant to demonstrate.

In our experience developing case studies, we found that identifying a data set first was often easier than relying on finding a data set to answer a particular question. While many of our case studies were specifically designed to address a public health challenge, we sometimes struggled to find publicly available data that was appropriate for the question or set of questions of interest. Collaboration with subject-matter experts can be especially helpful in addressing this challenge. For our case studies, we worked with public health experts in order to both identify interesting, timely, and testable questions and to find a public source of data to answer our questions. 

We found we could use the difficulty of obtaining data in a standard format (e.g., Excel, CSV) as a teaching opportunity, and that being open-minded about the source of the data allowed us to demonstrate unconventional skills. For example, when we could not easily access the data stored in a table in a published report, we illustrated the data science skill of pulling data directly from a PDF. As future data scientists, our students need the skills to be flexible to access data that cannot simply be read in or imported as-is into R.

While we typically started developing each case study with a set of data science and statistical learning objectives in mind, there was sometimes a tension between finding a data set that would allow us to meet these specific objectives and allowing the data to guide the direction of the case study. We found that following opportunities presented by the data itself led us to give examples that were more authentic to a real-world data analysis situation. We recorded some of these challenges within the case studies themselves so students could better understand the process of finding the right data to answer a question of interest (and the potential need to refocus a question). The limitations section in particular provides some of the most useful material for class discussions about the types of questions the data can and cannot answer and how sometimes we must simplify our analysis to reflect the limitations of the data available to us.

 As educators working during a time of reflection and social change around issues of gender and race in research, we also took care to point out some historically overlooked aspects of our data sets. For example, collecting data with surveys that provide a limited number of options about ethnicity or race or racial and gender intersections, limits our ability to accurately capture the diversity of the population being studied. As an example, we refer the reader to the case study about youth disconnection (\href{https://www.opencasestudies.org/ocs-bp-youth-disconnection}{opencasestudies.org/ocs-bp-youth-disconnection}).

For some case studies, we focused on finding mostly clean and complete data to allow us to demonstrate certain concepts, like machine learning or how to create a dashboard. In these case studies where we knew that the analytical material was going to get quite intensive and lengthy, we specifically sought to find data sets that would allow us to jump right in with little difficulty in terms of gathering, cleaning, and importing the data. 

Our overall suggestions for starting a case study are: 
\begin{itemize}[nosep]
 
 \item \textbf{Be open-minded and flexible about data sources:} Unlike performing a real analysis where an analyst might choose to avoid complications in accessing the data (when the option is available to go with a data set that is easier to access), such complications can provide teaching opportunities to prepare students for cases where they will not have a simpler option available. 
 
 \item \textbf{Determine the level of flexibility based on the goals of the case study:} If the case study is intended to demonstrate a specific statistical method or data import method, more effort may be required to find the right data to meet this specific teaching expectation. In our case we knew we were planning to make several case studies, thus we were able to let some of the case studies naturally flow in directions we didn't initially intend. This ultimately led to some teaching opportunities we did not expect. However, for some of our case studies we were more rigid about our data needs.
 
 \item \textbf{Think about the scope of the case study:} Keep in mind the 1) type of learners that the case study is intended for and 2) data analysis method goals that the case study is intended to demonstrate. Try to avoid a case study that is both intensive for data import/wrangling and intensive for data analysis. At a later point reevaluation of the overall direction and scope of the case study  may be needed. If the case study is too long, consider splitting it into multiple case studies.
 
 \item \textbf{Keep it simple:} Explaining a process at a beginner level often involved more space within a case study than anticipated. Keeping case study plans simple can help as unexpected teaching opportunities may arise that may require more instruction.
 
\end{itemize}

\subsection{Do the analysis first but with a learner in mind}

To present an analysis narrative, it is necessary to first perform the analysis before working on the narrative description. However, the case study itself should not simply be a reproduction of the process used to analyze the data.  Instead it should contain simplifications and modifications to create a clear and coherent presentation for students.  To do this, it is crucial to keep a good record of all the steps taken during this initial analysis, including explanations and comments to justify the analysis choices made along the way. Special care should be taken to record exactly how the raw data is obtained.

Often the way we would typically perform an analysis ourselves might not always be the best for instruction purposes. For example, an experienced data analyst might start by writing a function that wrangles multiple similar data files. However, this would not be the appropriate way to start a case study for beginners.  Instead one might choose to focus on wrangling a single specific file in great detail before trying to generalize the code as a function. Thus we try to determine an overall process of data import and wrangling for the intended level of audience before really generating the dialogue that describes this process. 

We also found that often the data exploration steps and the steps involved in the decision-making process of how to wrangle the data needed to be simplified for a case study. For example, we may ultimately decide to remove a data source from our analysis because we find errors in the data and dealing with these errors are beyond the scope for our intended audience. While it may be useful to tell students about these data errors \cite{peng_diagnosing_2021} and how to address them, we also need to keep an appropriate level of detail so as not to overwhelm them. 

Another situation where we might modify the analysis is if a process requires a considerable level of trial and error. Rather than showing the students all the iterations of the trial and error and all of the decisions around this process, we may only demonstrate a small portion so as not to make the case study too lengthy. In a case study about machine learning, for example, we aimed to achieve a certain level of performance so we spent a fair amount of time  demonstrating how to optimize and tune parameters. While we briefly described our tuning strategy, we did not show all intermediate models, but ultimately showed two that were interesting and useful for describing parameter tuning.

To conclude, we may have gone through a learning process in our own analysis, eventually arriving at a more refined approach. Instead of describing the entire process to get to this point, we would sometimes simply present the final approach, yet describe in the narrative that in practice more effort would be required. While we do want to present a realistic depiction of the data analysis process, we also need to achieve clarity and focus.

\subsection{Creating the case study narrative}

Once an analyst has performed the analysis to address the questions of interest, it is time to start writing the narrative. First, we introduce and motivate the main topic by presenting some research related to the particular question evaluated in the case study.

First, we describe the data import, wrangling, and analysis processes. As mentioned above, this will likely not be a faithful reproduction of our own analysis process, but will be recreated to best meet the pedagogical goals of the case study. In terms of added narrative, we do our best to guide students through the new information we are presenting. The first time we use a function we describe what it does, its main arguments, and what packages it comes from. We describe the thoughts behind our decision making process from one step to the next, sometimes illustrating times where we try something and it does not work to reflect a real-world data analysis. 

We also describe jargon and background information where possible with click-to-expand sections so as not to disrupt the general flow of the case study. For example, an expanded section would explain how "piping" works, passing objects through a series of steps, to avoid slowing down students who are already familiar, while allowing us to not lose students that have never seen piping before. Other material for such expandable sections includes describing the ``grammar of graphics" for the \texttt{ggplot2} package or providing background statistical information before performing a statistical test. In some cases we describe a concept at great length in another case study so we link to the description there, but in general we at least minimally describe most concepts and methods in each case study to keep them as self-contained as possible.  Similarly, we found including portions of RStudio cheat sheets \cite{rstudiocheatsheets} to be very useful for certain topics, such as describing regular expressions or joining functions. In some cases we found it best to explain a concept or challenge with a simpler example first using a smaller data set imported into R or created in R ourselves. This material is also included in click-to-expand sections for students who might already be familiar with such concepts. 

While constructing the narrative, we think about where we can include question opportunities. These opportunities include places for an instructor to start a discussion about the analysis decision-making process, such as why a particular graph choice is not always effective or why a wrangling method might not be reproducible. We may prompt students to try to remember how to perform a task that has already been shown in the case study previously. In our interactive case studies we also include quiz questions and coding exercises, as described in the Section~\ref{sec:ocs-resource-interactive}.

Finally, we end the narrative by summarizing how to communicate the major findings of the analysis \cite{khachatryan_v_2017}. We also describe how the results fit into the greater context of the field, what the implications are, the limitations of the study, and what is still unknown. We finish by going through the case study to create a list of all the resources shown throughout the case study. 

Through the process of creating this resource, we discovered a variety of challenges, as well as strategies that we used to overcome these challenges, as described in \textbf{Table~\ref{tab-challenges}} and guidelines for creating new case studies (\ref{sec-supp-note-guidelines}).

\subsection{Creating interactive case studies}
\label{sec:create-int-ocs}

 We have also included interactive elements in a subset of our case studies using packages (\texttt{learnr} \cite{learnr} and \texttt{gradethis} \cite{gradethis}) that build on the \texttt{shiny} \cite{shiny} package which allows 
 R users to more easily create web applications. The \texttt{learnr} package allows users to create multiple choice questions and coding exercises, while \texttt{gradethis} allows for customization of the feedback divided to learners as they answer questions or perform exercises. 
 
 There are two methods to do this. One method is to host each exercise as an individual Shiny application and then embed these applications in the case study using inline frames (HTML `iframe`). The second method is to create one single application that incorporates exercises within the case study (\ref{sec-supp-note-technical} and \ref{sec-supp-note-interactive}).

\section{Discussion and Conclusions}
\label{sec:discussion}

In this paper, we introduce a model for creating fully open-source, peer-reviewed, and complete case studies to create an archive of examples of best practices to guide students through data analyses involving real, complicated, messy, and interesting data. Our archive can be used in the classroom by instructors to guide students through any part of our case studies due to the easy navigation and common modularized architecture to structure the case studies. These can also be used by independent learners due to the thorough narrative, interactive elements, and complete analyses. Students and learners can learn about new topics or return to a case study to brush up on details of a particular method or technique. The data within our case studies and the narrated data analyses and data science methods can be used by instructors educating undergraduate and graduate students, as well as high school students in a variety of topics including statistics, public health, programming, and data science. This provides an opportunity for instructors to use data that is relevant to current public health concerns and therefore of interest to a large variety of students without the work required to identify such data or to determine what analyses are possible with such data. This will free instructors to focus on challenging the students with more interactive discussions in class and allow students to learn more about the decision processes required for analyzing data. 

In summary, OCS provides a consistent framework grounded by \cite{nolan1999}, is open, and additionally provides recommendations on how to teach the material. With the OCS resources, educators can also make their own and expand OCS if they contribute back. We believe these additions try to bridge the gaps in the last mile of analysis education.

\section*{Back Matter}

\subsection*{Author Contributions}
Contributions listed according to the CRediT system.

\begin{itemize}[nosep]
    \item Conceptualization: CW, MAT, LRJ, SCH
    \item Software: all co-authors
    \item Formal analysis: all co-authors
    \item Investigation: CW, SCH
    \item Data Curation	CW, SCH
    \item Writing - Original Draft: CW, SCH
    \item Writing - Review \& Editing: all co-authors
    \item Visualization: CW, MB
    \item Supervision: CW, MAT, LRJ, SCH
    \item Funding acquisition: CW, MAT, LRJ, SCH
\end{itemize}

\subsection*{Acknowledgements}

We would like to thank the Johns Hopkins Data Science lab (\href{https://jhudatascience.org/}{jhudatascience.org}), in particular Roger Peng, Jeff Leek, Brian Caffo, and Jessica Crowell for their support and valuable feedback on the Open Case Studies project. We would like to thank Ira Gooding for his feedback on incorproating case studies into the Coursera platform. In addition we would like to thank all the data science and statistics reviewers of our case studies, including: Shannon Ellis, Nicholas Horton, Leslie Myint, Mine Çetinkaya-Rundel, Michael Love, and Christina P. Knudson, as well as the following student reviewers: Jensen Stanton, Tina Trinh, and Ruby Ho. We would also like to acknowledge the topic reviewers including: Roger Peng, Tamar Mendelson, Brendan Saloner, Renee Johnson, Jessica Fanzo, Daniel Webster, Elizabeth Stuart, Aboozar Hadavand, Megan Latshaw, Kirsten Koehler, and Alexander McCourt. We would also like to acknowledge  Ashkan Afshin and Erin Mullany for giving us access to the data for the case study titled "Exploring global patterns of dietary behaviors associated with health risk." We would also like to thank the Johns Hopkins Bloomberg School of Public Health Department of Biostatistics for initially funding this project. 

\subsection*{Funding} 

The Open Case Study project reported in this publication was supported by a High-Impact Project grant in 2019-2020 by the Bloomberg American Health Initiative to create the majority of the case studies currently part of the project. A 2020 Digital Education \& Learning Technology Acceleration (DELTA) Grant from the Office of the Provost at the Johns Hopkins University supported the creation of interactive case studies and many of the tools that support their use, such as the search tool. The Open Case Studies guide was funded as an extension to funding for the Genomic Data Science Community Network (GDSCN). The GDSCN is supported through a contract to Johns Hopkins University (75N92020P00235) NHGRI. JM was supported by Streamline Data Science, U24HG010263-01 (ANVIL), UL1TR003098 (NIH/NCATS): Institutional Clinical and Translational Science Award, and UE5CA254170. 

\subsection*{Conflict of Interest} 

The co-authors Carrie Wright and Stephanie Hicks receive royalties on a book available on Leanpub and a course on Coursera titled ``Tidyverse Skills for Data Science'', both which incorporate three case studies from the Open Case Studies project.

\bibliographystyle{unsrtnat}
\bibliography{ocs-manuscript.bib}

\begin{thebibliography}{51}
\providecommand{\natexlab}[1]{#1}
\providecommand{\url}[1]{\texttt{#1}}
\expandafter\ifx\csname urlstyle\endcsname\relax
  \providecommand{\doi}[1]{doi: #1}\else
  \providecommand{\doi}{doi: \begingroup \urlstyle{rm}\Url}\fi

\bibitem[Wild and Pfannkuch(1999)]{wild1999}
C.~J. Wild and M.~Pfannkuch.
\newblock Statistical thinking in empirical enquiry.
\newblock \emph{International Statistical Review}, 67\penalty0 (3):\penalty0
  223--248, 1999.

\bibitem[Nolan and Speed(1999)]{nolan1999}
Deborah Nolan and Terry~P. Speed.
\newblock Teaching statistics theory through applications.
\newblock \emph{The American Statistician}, 53:\penalty0 370, 1999.

\bibitem[Hicks and Irizarry(2018)]{hicks2018}
Stephanie~C. Hicks and Rafael~A. Irizarry.
\newblock A guide to teaching data science.
\newblock \emph{The American Statistician}, 72\penalty0 (4):\penalty0 382--391,
  2018.
\newblock PMID: 31105314.

\bibitem[Weinberg and Abramowitz(2000)]{weinberg_casestudies_2000}
Sharon~Lawner Weinberg and Sarah~Knapp Abramowitz.
\newblock Making general principles come alive in the classroom using an active
  case studies approach.
\newblock \emph{Journal of Statistics Education}, 8\penalty0 (2):\penalty0
  null, 2000.
\newblock \doi{10.1080/10691898.2000.12131290}.
\newblock URL \url{https://doi.org/10.1080/10691898.2000.12131290}.

\bibitem[Schafer and Ramsey(2003)]{schafer_casestudies_2003}
Daniel~W. Schafer and Fred~L. Ramsey.
\newblock Teaching the craft of data analysis.
\newblock \emph{Journal of Statistics Education}, 11\penalty0 (1):\penalty0
  null, 2003.
\newblock \doi{10.1080/10691898.2003.11910692}.

\bibitem[Khachatryan and Karst(2017)]{khachatryan_v_2017}
Davit Khachatryan and Nathaniel Karst.
\newblock V for voice: Strategies for bolstering communication skills in
  statistics.
\newblock \emph{Journal of Statistics Education}, 25\penalty0 (2):\penalty0
  68--78, 2017.
\newblock ISSN 1069-1898.
\newblock \doi{10.1080/10691898.2017.1305261}.
\newblock URL
  \url{https://www.tandfonline.com/doi/full/10.1080/10691898.2017.1305261}.

\bibitem[Rivera et~al.(2019)Rivera, Marazzi, and
  Torres-Saavedra]{rivera_incorporating_2019}
Roberto Rivera, Mario Marazzi, and Pedro~A. Torres-Saavedra.
\newblock Incorporating open data into introductory courses in statistics.
\newblock \emph{Journal of Statistics Education}, 27\penalty0 (3):\penalty0
  198--207, 2019.
\newblock ISSN 1069-1898.
\newblock \doi{10.1080/10691898.2019.1669506}.
\newblock URL
  \url{https://www.tandfonline.com/doi/full/10.1080/10691898.2019.1669506}.

\bibitem[Donoghue et~al.(2021)Donoghue, Voytek, and
  Ellis]{donoghue_casestudies_2021}
Thomas Donoghue, Bradley Voytek, and Shannon~E. Ellis.
\newblock Teaching creative and practical data science at scale.
\newblock \emph{Journal of Statistics and Data Science Education}, 29\penalty0
  (sup1):\penalty0 S27--S39, 2021.
\newblock \doi{10.1080/10691898.2020.1860725}.

\bibitem[Reyes and McTavish(2022)]{reyes_casestudies_2022}
Luna L.~Sanchez Reyes and Emily~Jane McTavish.
\newblock Approachable case studies support learning and reproducibility in
  data science: An example from evolutionary biology.
\newblock \emph{Journal of Statistics and Data Science Education}, 30\penalty0
  (3):\penalty0 304--310, 2022.
\newblock \doi{10.1080/26939169.2022.2099487}.

\bibitem[Romero et~al.(1995)Romero, Ferrer, Capilla, Zunica, Balasch, Serra,
  and Alcover]{romero_teaching_1995}
R.~Romero, A.~Ferrer, C.~Capilla, L.~Zunica, S.~Balasch, V.~Serra, and
  R.~Alcover.
\newblock Teaching statistics to engineers: An innovative pedagogical
  experience.
\newblock \emph{Journal of Statistics Education}, 3\penalty0 (1):\penalty0 5,
  1995.
\newblock ISSN 1069-1898.
\newblock \doi{10.1080/10691898.1995.11910481}.
\newblock URL
  \url{https://www.tandfonline.com/doi/full/10.1080/10691898.1995.11910481}.

\bibitem[Theobold et~al.(2021)Theobold, Hancock, and
  Mannheimer]{theobold_casestudies_2021}
Allison~S. Theobold, Stacey~A. Hancock, and Sara Mannheimer.
\newblock Designing data science workshops for data-intensive environmental
  science research.
\newblock \emph{Journal of Statistics and Data Science Education}, 29\penalty0
  (sup1):\penalty0 S83--S94, 2021.
\newblock \doi{10.1080/10691898.2020.1854636}.

\bibitem[Arnold and Franklin(2021)]{arnold_what_2021}
Pip Arnold and Christine Franklin.
\newblock What makes a good statistical question?
\newblock \emph{Journal of Statistics and Data Science Education}, 29\penalty0
  (1):\penalty0 122--130, 2021.
\newblock ISSN 2693-9169.
\newblock \doi{10.1080/26939169.2021.1877582}.
\newblock URL
  \url{https://www.tandfonline.com/doi/full/10.1080/26939169.2021.1877582}.

\bibitem[Neumann et~al.(2013)Neumann, Hood, and Neumann]{neumann_using_2013}
David~L. Neumann, Michelle Hood, and Michelle~M. Neumann.
\newblock {USING} {REAL}-{LIFE} {DATA} {WHEN} {TEACHING} {STATISTICS}:
  {STUDENT} {PERCEPTIONS} {OF} {THIS} {STRATEGY} {IN} {AN} {INTRODUCTORY}
  {STATISTICS} {COURSE}.
\newblock \emph{{STATISTICS} {EDUCATION} {RESEARCH} {JOURNAL}}, 12\penalty0
  (2):\penalty0 59--70, 2013.
\newblock ISSN 1570-1824.
\newblock \doi{10.52041/serj.v12i2.304}.
\newblock URL \url{https://iase-web.org/ojs/SERJ/article/view/304}.

\bibitem[Donoho(2017)]{Donoho2017}
David Donoho.
\newblock 50 years of data science.
\newblock \emph{Journal of Computational and Graphical Statistics}, 26\penalty0
  (4):\penalty0 745--766, 2017.
\newblock \doi{10.1080/10618600.2017.1384734}.
\newblock URL \url{https://doi.org/10.1080/10618600.2017.1384734}.

\bibitem[Wood et~al.(2018)Wood, Mocko, Everson, Horton, and
  Velleman]{Wood_2018}
Beverly~L. Wood, Megan Mocko, Michelle Everson, Nicholas~J. Horton, and Paul
  Velleman.
\newblock Updated guidelines, updated curriculum: The {GAISE} college report
  and introductory statistics for the modern student.
\newblock \emph{{CHANCE}}, 31\penalty0 (2):\penalty0 53--59, April 2018.
\newblock \doi{10.1080/09332480.2018.1467642}.
\newblock URL
  \url{https://www.tandfonline.com/doi/full/10.1080/09332480.2018.1467642}.

\bibitem[{Committee on Envisioning the Data Science Discipline: The
  Undergraduate Perspective} et~al.(2018){Committee on Envisioning the Data
  Science Discipline: The Undergraduate Perspective}, {Computer Science and
  Telecommunications Board}, {Board on Mathematical Sciences and Analytics},
  {Committee on Applied and Theoretical Statistics}, {Division on Engineering
  and Physical Sciences}, {Board on Science Education}, {Division of Behavioral
  and Social Sciences and Education}, and {National Academies of Sciences,
  Engineering, and
  Medicine}]{committee_on_envisioning_the_data_science_discipline_the_undergraduate_perspective_data_2018}
{Committee on Envisioning the Data Science Discipline: The Undergraduate
  Perspective}, {Computer Science and Telecommunications Board}, {Board on
  Mathematical Sciences and Analytics}, {Committee on Applied and Theoretical
  Statistics}, {Division on Engineering and Physical Sciences}, {Board on
  Science Education}, {Division of Behavioral and Social Sciences and
  Education}, and {National Academies of Sciences, Engineering, and Medicine}.
\newblock \emph{Data Science for Undergraduates: Opportunities and Options}.
\newblock National Academies Press, 2018.
\newblock ISBN 9780309475594.
\newblock \doi{10.17226/25104}.
\newblock URL \url{https://www.nap.edu/catalog/25104}.

\bibitem[Kross and Guo(2019)]{KrossGuo2019}
Sean Kross and Philip~J. Guo.
\newblock Practitioners teaching data science in industry and academia:
  Expectations, workflows, and challenges.
\newblock In \emph{Proceedings of the 2019 CHI Conference on Human Factors in
  Computing Systems}, CHI '19, page 1–14, New York, NY, USA, 2019.
  Association for Computing Machinery.
\newblock ISBN 9781450359702.

\bibitem[Waller(2018)]{Waller2018}
Lance~A. Waller.
\newblock Documenting and evaluating data science contributions in academic
  promotion in departments of statistics and biostatistics.
\newblock \emph{The American Statistician}, 72\penalty0 (1):\penalty0 11--19,
  2018.
\newblock \doi{10.1080/00031305.2017.1375988}.

\bibitem[Peng et~al.(2021)Peng, Chen, Bridgeford, Leek, and
  Hicks]{peng_diagnosing_2021}
Roger~D. Peng, Athena Chen, Eric Bridgeford, Jeffrey~T. Leek, and Stephanie~C.
  Hicks.
\newblock Diagnosing data analytic problems in the classroom.
\newblock \emph{Journal of Statistics and Data Science Education}, 29\penalty0
  (3):\penalty0 267--276, 2021.
\newblock ISSN 2693-9169.
\newblock \doi{10.1080/26939169.2021.1971586}.
\newblock URL
  \url{https://www.tandfonline.com/doi/full/10.1080/26939169.2021.1971586}.

\bibitem[Vilhuber et~al.(2022)Vilhuber, Son, Welch, Wasser, and
  Darisse]{vilhuber_teaching_2022}
Lars Vilhuber, Hyuk~Harry Son, Meredith Welch, David~N. Wasser, and Michael
  Darisse.
\newblock Teaching for large-scale reproducibility verification.
\newblock \emph{Journal of Statistics and Data Science Education}, 30\penalty0
  (3):\penalty0 274--281, 2022.
\newblock ISSN 2693-9169.
\newblock \doi{10.1080/26939169.2022.2074582}.
\newblock URL
  \url{https://www.tandfonline.com/doi/full/10.1080/26939169.2022.2074582}.

\bibitem[Dogucu and Çetinkaya Rundel(2022)]{dogucu_tools_2022}
Mine Dogucu and Mine Çetinkaya Rundel.
\newblock Tools and recommendations for reproducible teaching.
\newblock \emph{Journal of Statistics and Data Science Education}, 30\penalty0
  (3):\penalty0 251--260, 2022.
\newblock ISSN 2693-9169.
\newblock \doi{10.1080/26939169.2022.2138645}.
\newblock URL
  \url{https://www.tandfonline.com/doi/full/10.1080/26939169.2022.2138645}.

\bibitem[Nolan and Lang(2010)]{nolan_computing}
Deborah Nolan and Duncan~Temple Lang.
\newblock Computing in the statistics curricula.
\newblock \emph{The American Statistician}, 64\penalty0 (2):\penalty0 97--107,
  2010.
\newblock \doi{10.1198/tast.2010.09132}.

\bibitem[{R Core Team}(2021)]{R_2021}
{R Core Team}.
\newblock \emph{R: A Language and Environment for Statistical Computing}.
\newblock R Foundation for Statistical Computing, Vienna, Austria, 2021.
\newblock URL \url{https://www.R-project.org/}.

\bibitem[Çetinkaya Rundel(2018)]{mine2018}
Mine Çetinkaya Rundel.
\newblock \emph{{Let Them Eat Cake (First)! [Video]. YouTube. (2018, November
  9)}}, 2018.
\newblock URL \url{https://www.youtube.com/watch?v=RsVOrpXAPXo}.

\bibitem[Ratan et~al.(2019)Ratan, Anand, and Ratan]{ratan_formulation_2019}
Simmi~K. Ratan, Tanu Anand, and John Ratan.
\newblock Formulation of {Research} {Question} – {Stepwise} {Approach}.
\newblock \emph{Journal of Indian Association of Pediatric Surgeons},
  24\penalty0 (1):\penalty0 15, March 2019.
\newblock URL \url{https://www.ncbi.nlm.nih.gov/pmc/articles/PMC6322175/}.
\newblock Publisher: Wolters Kluwer -- Medknow Publications.

\bibitem[Osueke et~al.(2018)Osueke, Mekonnen, and Stanton]{osueke_how_2018}
Bethany Osueke, Birook Mekonnen, and Julie~Dangremond Stanton.
\newblock How {Undergraduate} {Science} {Students} {Use} {Learning}
  {Objectives} to {Study}.
\newblock \emph{Journal of Microbiology \& Biology Education}, 19\penalty0 (2),
  January 2018.
\newblock ISSN 1935-7877, 1935-7885.
\newblock URL \url{https://journals.asm.org/doi/10.1128/jmbe.v19i2.1510}.

\bibitem[Wickham et~al.(2022)Wickham, François, Lionel, and
  Müller]{pkg-dplyr}
Hadley Wickham, Romain François, Henry Lionel, and Kirill Müller.
\newblock \emph{{dplyr: A Grammar of Data Manipulation}}, 2022.
\newblock URL \url{https://CRAN.R-project.org/package=dplyr}.

\bibitem[Wickham(2016)]{ggplot2}
Hadley Wickham.
\newblock \emph{ggplot2: Elegant Graphics for Data Analysis}.
\newblock Springer-Verlag New York, 2016.
\newblock ISBN 978-3-319-24277-4.
\newblock URL \url{https://ggplot2.tidyverse.org}.

\bibitem[Knuth(1984)]{knuth_literate_1984}
D.~E. Knuth.
\newblock Literate {Programming}.
\newblock \emph{The Computer Journal}, 27\penalty0 (2):\penalty0 97--111,
  January 1984.
\newblock ISSN 0010-4620.
\newblock \doi{10.1093/comjnl/27.2.97}.
\newblock URL \url{https://doi.org/10.1093/comjnl/27.2.97}.

\bibitem[Wickham and Hester(2020)]{pkg-readr}
Hadley Wickham and Jim Hester.
\newblock \emph{{readr: Read Rectangular Text Data}}, 2020.
\newblock URL \url{https://CRAN.R-project.org/package=readr}.

\bibitem[Tukey(1962)]{Tukey1962}
John~W. Tukey.
\newblock {The Future of Data Analysis}.
\newblock \emph{The Annals of Mathematical Statistics}, 33\penalty0
  (1):\penalty0 1 -- 67, 1962.

\bibitem[Müller(2020)]{here}
Kirill Müller.
\newblock \emph{here: A Simpler Way to Find Your Files}, 2020.
\newblock URL \url{https://CRAN.R-project.org/package=here}.
\newblock R package version 1.0.1.

\bibitem[Wickham and Bryan(2022)]{readxl}
Hadley Wickham and Jennifer Bryan.
\newblock \emph{readxl: Read Excel Files}, 2022.
\newblock URL \url{https://CRAN.R-project.org/package=readxl}.
\newblock R package version 1.4.0.

\bibitem[Bache and Wickham(2022)]{magrittr}
Stefan~Milton Bache and Hadley Wickham.
\newblock \emph{magrittr: A Forward-Pipe Operator for R}, 2022.
\newblock URL \url{https://CRAN.R-project.org/package=magrittr}.
\newblock R package version 2.0.3.

\bibitem[Wickham(2022)]{stringr}
Hadley Wickham.
\newblock \emph{stringr: Simple, Consistent Wrappers for Common String
  Operations}, 2022.
\newblock URL \url{https://CRAN.R-project.org/package=stringr}.
\newblock R package version 1.4.1.

\bibitem[Henry and Wickham(2020)]{purrr}
Lionel Henry and Hadley Wickham.
\newblock \emph{purrr: Functional Programming Tools}, 2020.
\newblock URL \url{https://CRAN.R-project.org/package=purrr}.
\newblock R package version 0.3.4.

\bibitem[Wickham and Girlich(2022)]{tidyr}
Hadley Wickham and Maximilian Girlich.
\newblock \emph{tidyr: Tidy Messy Data}, 2022.
\newblock URL \url{https://CRAN.R-project.org/package=tidyr}.
\newblock R package version 1.2.0.

\bibitem[Müller and Wickham(2022)]{tibble}
Kirill Müller and Hadley Wickham.
\newblock \emph{tibble: Simple Data Frames}, 2022.
\newblock URL \url{https://CRAN.R-project.org/package=tibble}.
\newblock R package version 3.1.8.

\bibitem[Wickham(2021)]{forcats}
Hadley Wickham.
\newblock \emph{forcats: Tools for Working with Categorical Variables
  (Factors)}, 2021.
\newblock URL \url{https://CRAN.R-project.org/package=forcats}.
\newblock R package version 0.5.1.

\bibitem[Hocking(2021)]{directlabels}
Toby~Dylan Hocking.
\newblock \emph{directlabels: Direct Labels for Multicolor Plots}, 2021.
\newblock URL \url{https://CRAN.R-project.org/package=directlabels}.
\newblock R package version 2021.1.13.

\bibitem[Slowikowski(2021)]{ggrepel}
Kamil Slowikowski.
\newblock \emph{ggrepel: Automatically Position Non-Overlapping Text Labels
  with 'ggplot2'}, 2021.
\newblock URL \url{https://CRAN.R-project.org/package=ggrepel}.
\newblock R package version 0.9.1.

\bibitem[Robinson et~al.(2022)Robinson, Hayes, and Couch]{broom}
David Robinson, Alex Hayes, and Simon Couch.
\newblock \emph{broom: Convert Statistical Objects into Tidy Tibbles}, 2022.
\newblock URL \url{https://CRAN.R-project.org/package=broom}.
\newblock R package version 1.0.0.

\bibitem[Pedersen(2020)]{patchwork}
Thomas~Lin Pedersen.
\newblock \emph{patchwork: The Composer of Plots}, 2020.
\newblock URL \url{https://CRAN.R-project.org/package=patchwork}.
\newblock R package version 1.1.1.

\bibitem[Xie et~al.(2021)Xie, Cheng, and Tan]{DT}
Yihui Xie, Joe Cheng, and Xianying Tan.
\newblock \emph{DT: A Wrapper of the JavaScript Library 'DataTables'}, 2021.
\newblock URL \url{https://CRAN.R-project.org/package=DT}.
\newblock R package version 0.17.

\bibitem[noa(2021)]{noauthorfeaturesnodate}
\emph{{GitHub} {Actions}}, 2021.
\newblock URL \url{https://github.com/features/actions}.

\bibitem[Shahin et~al.(2017)Shahin, Ali~Babar, and Zhu]{shahin_continuous_2017}
Mojtaba Shahin, Muhammad Ali~Babar, and Liming Zhu.
\newblock Continuous {Integration}, {Delivery} and {Deployment}: {A}
  {Systematic} {Review} on {Approaches}, {Tools}, {Challenges} and {Practices}.
\newblock \emph{IEEE Access}, 5:\penalty0 3909--3943, 2017.
\newblock ISSN 2169-3536.
\newblock URL \url{http://ieeexplore.ieee.org/document/7884954/}.

\bibitem[Schloerke et~al.(2020)Schloerke, Allaire, and Borges]{learnr}
Barret Schloerke, JJ~Allaire, and Barbara Borges.
\newblock \emph{learnr: Interactive Tutorials for R}, 2020.
\newblock URL \url{https://CRAN.R-project.org/package=learnr}.
\newblock R package version 0.10.1.

\bibitem[Aden-Buie et~al.(2021)Aden-Buie, Chen, Grolemund, and
  Schloerke]{gradethis}
Garrick Aden-Buie, Daniel Chen, Garrett Grolemund, and Barret Schloerke.
\newblock \emph{gradethis: Automated Feedback for Student Exercises in 'learnr'
  Tutorials}, 2021.
\newblock URL \url{https://pkgs.rstudio.com/gradethis/}.

\bibitem[rst(2022)]{rstudiocheatsheets}
\emph{{RStudio} {Cheatsheets}}, 2022.
\newblock URL \url{https://rstudio.com/resources/cheatsheets/}.

\bibitem[Chang et~al.(2021)Chang, Cheng, Allaire, Sievert, Schloerke, Xie,
  Allen, McPherson, Dipert, and Borges]{shiny}
Winston Chang, Joe Cheng, JJ~Allaire, Carson Sievert, Barret Schloerke, Yihui
  Xie, Jeff Allen, Jonathan McPherson, Alan Dipert, and Barbara Borges.
\newblock \emph{shiny: Web Application Framework for R}, 2021.
\newblock URL \url{https://CRAN.R-project.org/package=shiny}.
\newblock R package version 1.6.0.

\bibitem[De~Leon(2022)]{de_leon_interactive}
Desirée De~Leon.
\newblock \emph{Interactive tutorial windows for your {R} {Markdown} {Site}},
  2022.
\newblock URL \url{https://desiree.rbind.io/blog/learnr-iframes/}.

\end{thebibliography}

\clearpage
\onecolumn

\rfoot{Wright et al.\hspace{7pt}$\mid$\hspace{7pt}2023\hspace{7pt}$\mid$\hspace{7pt}ar\textcolor{black}{X}$\chi$iv\hspace{7pt}$\mid$\hspace{7pt} Page S\thepage}

{\huge Supplementary Materials}

\hrule

\vspace*{0.5cm}

\begin{center}

{\Large Open Case Studies: Statistics and Data Science Education through Real-World Applications}

\vspace*{0.75cm}

{\large Carrie Wright, Qier Meng, Michael Breshock, Margaret A. Taub, Leah R. Jager, John Muschelli, Stephanie C.\ Hicks$^*$}

\vspace*{0.3cm}

{\small $^*$Correspondence to \url{shicks19@jhu.edu}}

\end{center}

\renewcommand{\figurename}{Supplementary Figure}
\renewcommand{\tablename}{Supplementary Table}
\setcounter{figure}{0}
\setcounter{table}{0}
\setcounter{section}{0}
\setcounter{page}{1}
\makeatletter
\renewcommand{\thefigure}{S\@arabic\c@figure}
\renewcommand{\thetable}{S\@arabic\c@table}
\renewcommand{\thesection}{Supplemental Note S\@arabic\c@section}
\makeatother

\vspace*{1cm}

{\bf \large Contents}

\begin{enumerate}
    \item \textbf{Supplemental Table~S1-S3.}
    \item \textbf{Supplemental Figures~S1-S5.}
    \item \textbf{Supplemental Notes~S1-S3.}
\end{enumerate}

\clearpage

\noindent {\LARGE Supplemental Tables}

\begin{table*}[ht!]
\centering
\begin{framed}
\footnotesize
    \begin{tabular}{|c|L{6cm}L{6cm}|}
    \hline
    \multicolumn{3}{|c|}{\textbf{Diversity of topics in the Open Case Study resource}} \\ 
    \hline
    \cellcolor{Gray}Question  &  \cellcolor{Gray}How does something change over time? 
    \cellcolor{Gray} & \cellcolor{Gray}How do groups compare?\\
    \cellcolor{Gray}Types & \cellcolor{Gray}How do groups compare over time? &
    \cellcolor{Gray}How do paired groups compare?\\
    \cellcolor{Gray}&\cellcolor{Gray}Are certain groups or subgroups more vulnerable? &
    \cellcolor{Gray}How does something compare across regions?\\
    \cellcolor{Gray}& \cellcolor{Gray}How to predict outcomes for new data? &
    \cellcolor{Gray}Does this influence my data?\\
    \cellcolor{Gray}&\cellcolor{Gray}Are variables related to one another? &
    \cellcolor{Gray}How to display this data?\\
    \hline
    Data & Multiple files& PDF \\
    Types & CSV & Excel\\
    & Website & Image text \\
    & API & Google Sheets\\
    & Survey data / Codebooks & \\
    \hline
    \cellcolor{Gray}Wrangling  & \cellcolor{Gray}Extracting data from a PDF & 
    \cellcolor{Gray}Geocoding data \\
    \cellcolor{Gray}Methods & \cellcolor{Gray}Recoding data & 
    \cellcolor{Gray}Joining data \\
    \cellcolor{Gray} & \cellcolor{Gray}Modifying data &
    \cellcolor{Gray}Working with text\\
    \cellcolor{Gray}& \cellcolor{Gray}Reshaping data  & 
    \cellcolor{Gray}Repetitive process\\
    \hline
    Data & Formatted Table & Scatter plot \\
    Visualizations & Line plot & Bar plot \\
    & Box plots & Pie chart / Waffle plot \\
    & Heat map & Correlation plots \\
    & Missing data plots & Maps \\
    \hline
    \cellcolor{Gray} Advanced &  \cellcolor{Gray}Matching a plot style & 
    \cellcolor{Gray}Faceted plots\\
    \cellcolor{Gray} Visualizations & \cellcolor{Gray}Direct group labels & 
    \cellcolor{Gray}Emphasizing a group\\
    \cellcolor{Gray}& \cellcolor{Gray}Plot annotations & 
    \cellcolor{Gray}Plot error bars \\
    \cellcolor{Gray}& \cellcolor{Gray}Combining plots & 
    \cellcolor{Gray}Interactive plots \\
    \cellcolor{Gray}& \cellcolor{Gray}Interactive maps & 
    \cellcolor{Gray}Interactive tables \\
    \cellcolor{Gray}& \cellcolor{Gray}Adding images to plots & 
    \cellcolor{Gray}Interactive dashboard\\
    \hline
    Analysis & t-tests &ANOVA  \\
    Concepts & Linear Regression &  Logistic Regression\\
    and  &  Mann-Kendall Trend Test & Machine Learning \\
    Methods & Chi-Squared Test of Independence & Wilcoxon Signed-Rank Test\\
    & Calculating percentages with missing data & Wilcoxon Rank Sum Test\\
    \hline
    \end{tabular}
\caption{\textbf{Example topics in the OCS resource \wangbar} An example of how these are applied in an example case studies is provided in \textbf{Table~\ref{tab-exOCS}}.}
\label{tab-topics}
\end{framed}
\end{table*}

\begin{table*}[h!]
\scriptsize
\centering
\begin{framed}
\begin{tabular}{|C{4cm}|C{4cm}|C{4cm}|}
\hline
 \multicolumn{3}{|c|}{\textbf{Examples of modular case-study use}}\\
\hline
\cellcolor{DarkGray} User & \cellcolor{DarkGray} Example Use  & \cellcolor{DarkGray} Data Folder \\
\hline
\end{tabular}
\begin{tabular}{|C{4cm}|L{4cm}|C{4cm}|}
\cellcolor{Gray} Student & \cellcolor{Gray}Data science students looking for open source data for a class project &\cellcolor{Gray} raw\\
\hline
 Student & Public health student practicing data wrangling and visualization & imported\\
 \hline
\cellcolor{Gray} Educator & \cellcolor{Gray}Course instructor assigns homework using related but new data that expands beyond the case study &\cellcolor{Gray} extra\\
\hline
 Educator &  Data analysis instructor who wants students to practice some simple data import, but has limited time & simpler\textunderscore  import\\
\hline
\cellcolor{Gray} Self-Learner & \cellcolor{Gray}Researcher looking for analysis examples &\cellcolor{Gray} wrangled\\
\hline
\end{tabular}
\caption{\textbf{Examples of modular case-study use \wangbar} Examples of potential uses for case studies, beyond demonstrating the full case study in a class or as a learner following along.}
\label{tab-mod}
\end{framed}
\end{table*}

\clearpage
\begin{table*}[ht!]
\begin{framed}
\footnotesize
\centering
\begin{tabular}{|C{8cm}|C{8cm}|}
\hline
 \multicolumn{2}{|c|}{\textbf{Challenges of creating case studies in the OCS resource}}\\
\hline
\cellcolor{DarkGray} Challenge & \cellcolor{DarkGray} A Suggested Solution \\
\hline
\end{tabular}
\begin{tabular}{|L{8cm}|L{8cm}|}
\cellcolor{Gray}Finding the appropriate scope, data, and questions for the intended audience & \cellcolor{Gray}Be open minded about data sources, be flexible about revising or removing data\\
\hline
Balancing the teaching goals with the teaching opportunities presented by the data & Keeping the plan simple to allow room for unexpected teaching opportunities \\
\hline
\cellcolor{Gray}Showing the right amount of the data science process & \cellcolor{Gray}Performing the analysis first, then curating what will be included\\
\hline
Balancing an assumption of some prior knowledge with making the case study self-contained & Click-to-expand sections for additional information and links to other case studies and resources  \\
\hline
\cellcolor{Gray}Catering to multiple audience types  & \cellcolor{Gray}Modularizing case studies to allow different users to use only certain parts of the case studies and including click-to-expand sections for students that need more background\\
\hline
Modularization of case studies & Saving the data after each section and loading at the beginning of each section \\
\hline
\end{tabular}
\caption{\textbf{Challenges of creating case studies in the OCS resource \wangbar} A list of the major challenges that we experienced and suggested solutions.}
\label{tab-challenges}
\end{framed}
\end{table*}

\clearpage

\noindent {\LARGE Supplemental Figures}

\begin{figure*}[hb!]
\centering
\begin{framed}
\includegraphics[width=0.75\textwidth]{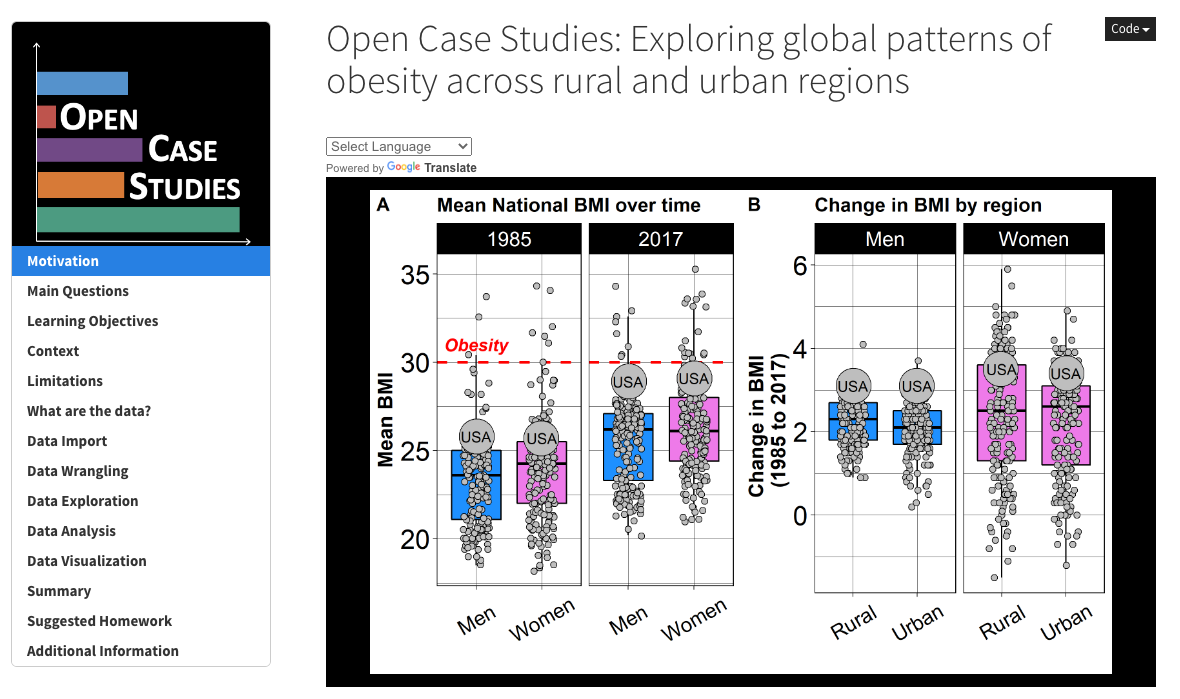}
\caption{
 {\small \textbf{Case Study Structure \wangbar} An image of the top of a case study showing the interactive table of contents to the left and a image summarizing the main findings of the case study.} 
 } 
\label{ocs_str}
\end{framed}
\end{figure*}

\begin{figure*}[hb!]
\centering
\begin{framed}
\includegraphics[width=0.80\textwidth]{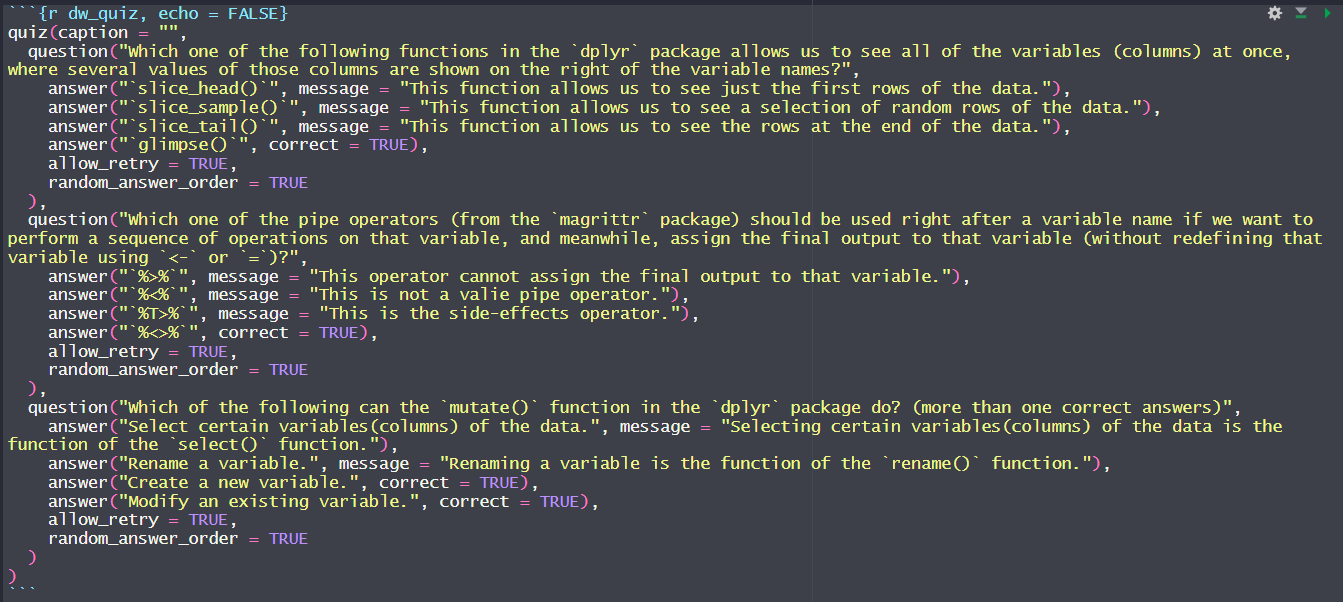}
 \caption{
 {\small \textbf{Creating a Multiple Choice Quiz with \texttt{learnr} \wangbar} An example code chunk used to create a multiple choice quiz in the \href{https://rsconnect.biostat.jhsph.edu/ocs-bp-co2-emissions-interactive/}{CO2 Emissions case study}.} 
 } 
  \label{learnr_quiz_code}
\end{framed}
\end{figure*}

\begin{figure*}[hb!]
\centering
\begin{framed}
\includegraphics[width=0.75\textwidth]{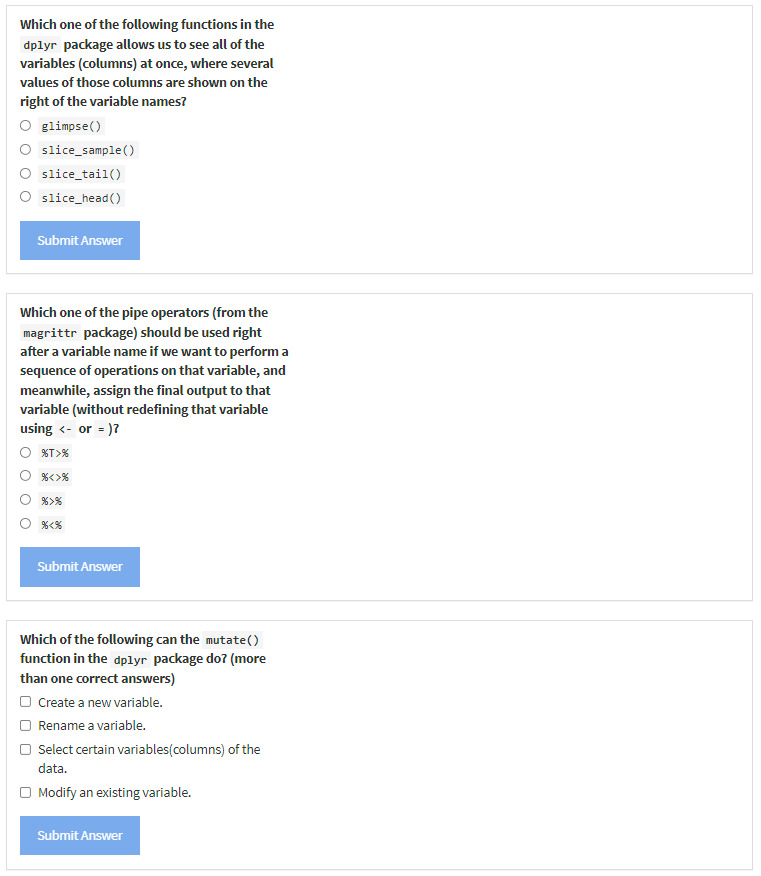}
 \caption{
 {\small \textbf{Multiple Choice Quiz Rendered in Case Study \wangbar} The multiple choice quiz in the \href{https://rsconnect.biostat.jhsph.edu/ocs-bp-co2-emissions-interactive/}{CO2 Emissions case study} created using the code in Figure \ref{learnr_quiz_code}.}
 } 
  \label{learnr_quiz_render}
\end{framed}
\end{figure*}

\begin{figure*}[hb!]
\centering
\begin{framed}
\includegraphics[width=0.95\textwidth]{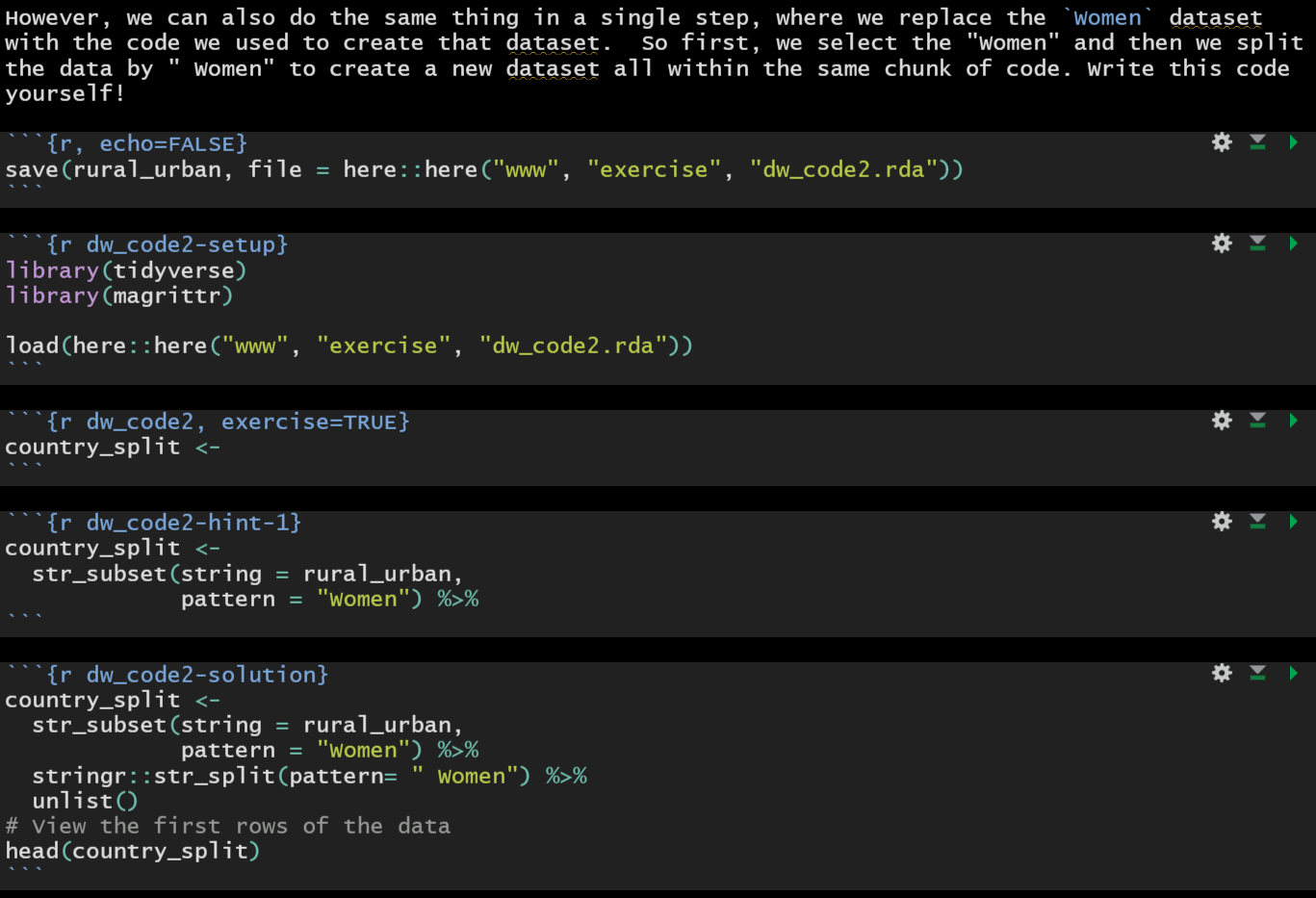}
 \caption{
 {\small \textbf{Creating a Coding Exercise with \texttt{learnr} \wangbar} An example of the code chunks used to create a coding exercise in the \href{https://rsconnect.biostat.jhsph.edu/ocs-bp-rural-and-urban-obesity-interactive/}{Obesity case study}.}
 } 
  \label{learnr_exercise_code}
\end{framed}
\end{figure*}

\begin{figure*}[hb!]
\centering
\begin{framed}
\includegraphics[width=0.95\textwidth]{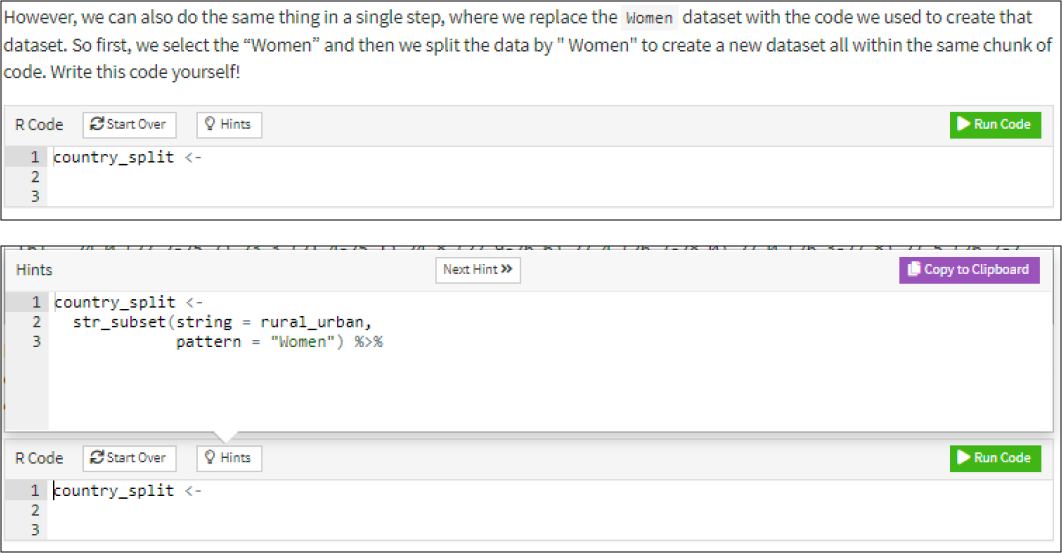}
 \caption{
 {\small \textbf{Coding Exercise Rendered in Case Study \wangbar} The coding exercise rendered in the \href{https://rsconnect.biostat.jhsph.edu/ocs-bp-rural-and-urban-obesity-interactive/}{Obesity case study} created using the code in Figure \ref{learnr_exercise_code}. Top: The exercise prompt. Bottom: Displays the hint function that can be used when stuck. The hints are made also with \texttt{learnr}.}
 } 
  \label{learnr_exercise_render}
\end{framed}
\end{figure*}

\clearpage

\noindent {\LARGE Supplemental Notes}

\makeatletter

\section{Guidelines for a creating case study in the OCS collection}
\label{sec-supp-note-guidelines}

To preserve the integrity of the core ideas of our resource we suggest the following guidelines for case studies to be included in the Open Case Studies collection:

\begin{itemize}[nosep]

 \item Case studies should be written in open source programming languages.
 \item Case studies should use data that is publicly available or can be made publicly available. Please ensure that the data can be made public if it is not already.
\item Case studies should include disclaimers and appropriate license agreements.
\item Effort should be made to describe the original source of the data with transparency.
\item All included images (that are not original to the case study) should include a source.
\item Core sections of the case study are required: Motivation, Main Questions, Learning Objectives, Limitations (outlining the limitations of the data), What are the Data?, and Summary (including limitations for the analysis presented).
\item Case studies should aim to describe the decision making process involved in performing data-science related tasks.
\item Links to literature or other sources to motivate the scientific topic of the case study should be included where possible.
\item Despite often being motivated by articles, case studies are not intended to demonstrate the methods of a paper. They are intended as an educational resource where users are guided through the data science process.

\end{itemize}

\section{Technical aspects of the two methods for creating interactive case studies}
\label{sec-supp-note-technical}

In this note we describe how we implemented the two options for creating interactive case studies: 1) hosting each exercise as an individual Shiny application and then embedding these applications in the case study or 2) creating one single application that incorporates exercises within the case study by creating the case study as a \texttt{learnr} tutorial.

\paragraph{Approach 1 - Embedding interactive elements.}
 
We did this by adapting the method of \citet{de_leon_interactive}. Building upon the iFrame Resizer JavaScript library, De Leon introduced the use of an HTML file that resizes the \texttt{learnr} tutorial windows in real time. This HTML file is included in the YAML header of the R Markdown file for the exercise, along with div tags in the last line to indicate the end of the content for the iFrame Resizer. We also added this feature by creating another HTML file containing the iFrame Resizer JavaScript and including it in the YAML header. Then, each exercise Shiny application is added to the case studies as an `iframe` with class “interactive”. This is done by creating the \texttt{learnr} exercises in separate R Markdown files. Each exercise is then published as an individual Shiny app online. Once published, the exercises now each have a unique URL address where they are hosted. The exercises can now be rendered within re-sizable windows inside the case study itself by nesting iFrame HTML code chunks within the case study R Markdown file using the exercise URLs as inputs.  At the end of each case study R Markdown, an HTML tag specifying the application of the resizer only to class “interactive” is added. R \texttt{knitr} has an \texttt{include\_app()} function to embed Shiny applications,  but De Leon's method better accommodates the aesthetics of the website as well as the pop-ups in the tutorials like hints, solutions, and feedback. 

\paragraph{Approach 2 - Single interactive case study application.}

The second option is to make the entire case study a \texttt{learnr} tutorial, where exercises can be directly created in the case study R Markdown document. To do this, we add \texttt{runtime: shiny\_prerendered} to the YAML header of this document. In order to publish the single case study application, all of the files needed to render the case study (data, images, CSS style code, etc.) need to be put in a \texttt{www/} sub-directory. This folder will be published along with the \texttt{index.Rmd} file to Shiny. The quizzes can be created inside a single R code chunk using the \texttt{quiz()}, \texttt{question()}, and \texttt{answer()} functions from \texttt{learnr} (\textbf{Figures \ref{learnr_quiz_code}-\ref{learnr_quiz_render}}). The coding exercises must be created using multiple R chunks. One chunk is written for each of the following: exercise setup, hints, solutions, and the exercise prompt (\textbf{Figure~\ref{learnr_exercise_code}}). In the rendered case study, these individual chunks are combined into a single coding exercise element (\textbf{Figure~\ref{learnr_exercise_render}}). Eventually, this document is published as a single Shiny application (\href{https://www.opencasestudies.org/ocs-make-interactive-tutorial/}{opencasestudies.org/ocs-make-interactive-tutorial}).

\paragraph{Differences between the approaches.}
This first approach of creating interactive elements is beneficial as it is generally easier to maintain. If one exercise breaks, we can readily identify which exercise is broken. However, the method can be inefficient as multiple Shiny applications are needed for a single case study and since each application stands alone, the data in the case study cannot be directly passed into the exercise. Thus, when the exercise involves using the data from the case study at a specific point in the analysis, the setup can become more complicated. 

The second approach is more efficient because the entire case study is written in a single file. However, the startup of a Shiny application is required to be under 60 seconds. For case studies that take longer to render, some of the data analysis results need to be stored separately to reduce the startup time. Also, since a single file is more difficult to maintain, it is important to keep detailed maintenance notes for the case study. 

We decided to go with the second approach, because we created many exercises and this would have resulted in tens of Shiny applications. The first approach involves making each exercise question a \texttt{learnr} tutorial, which is then published as a Shiny application or with Posit Connect. With this approach, the actual rendered version of the interactive case study is still hosted on GitHub, but each exercise is embedded in the website as its own window using HTML `iframe`. The second approach, that we we ultimately adopted, involves making the interactive case study a single \texttt{learnr} tutorial and in this case, instead of the rendered website version being hosted on GitHub, it is published as a Shiny application. In our case, we chose to use Posit Connect for this which does involve some payment. Currently, Shiny allows users to publish up to 5 applications for free for a single individual.

\section{Examples of of interactive questions in case studies}
\label{sec-supp-note-interactive}

We included a couple of types of interactive elements. The first are coding exercises as part of the analysis in the case study. These exercises are placed where the analysis repeats what is already done with another data set. The students apply what they learned, while staying in the context of the case study. The second type are multiple choice quiz questions and coding exercises at the end of case study sections.

As an example of the first type, if there are two similar data sets that need to have their variable names changed, we might demonstrate how to do it with one data set and ask the student to change the variable names of the other. Other circumstances for these exercises include, for instance, where an argument for a function is introduced. The exercise window can be used to explore the function of that argument by implementing the code with or without such argument. For example, when introducing the \texttt{set.seed()} function, the exercise window allows the students to experiment on how using a different seed would change the output.

The second type of interactive elements are to assist with learners assessing their knowledge using quiz questions and coding exercises at the end of case study sections. We include these exercises in a separate subsection so that they are easy to navigate to. Instead of focusing on a specific step of the analysis, these problems integrate the content of the entire section. The multiple-choice questions test the students’ understanding of the concepts introduced. For example, they may be asked what the key assumptions of a linear regression model are, what an R function can do, etc. The coding questions check the students’ abilities to implement the R functions in data analyses. For example, they could be asked to convert a made-up data set from wide format to long format, to build a machine learning model using an R built-in data set, etc. 

\makeatother

\end{document}